\documentclass[pdflatex,sn-mathphys]{sn-jnl}% Math and Physical Sciences Reference Style
\jyear{2021}%

\theoremstyle{thmstyleone}%
\theoremstyle{thmstyletwo}%

\theoremstyle{thmstylethree}%
\usepackage{makecell}

\usepackage{tabularx}
\usepackage{graphicx}
\usepackage{changepage}
\newcommand\norm[1]{\left\lVert#1\right\rVert}

\begin{document}

\title[Article Title]{Unlocking the capabilities of explainable few-shot learning in remote sensing}

%%=============================================================%%
%% Prefix	-> \pfx{Dr}
%% GivenName	-> \fnm{Joergen W.}
%% Particle	-> \spfx{van der} -> surname prefix
%% FamilyName	-> \sur{Ploeg}
%% Suffix	-> \sfx{IV}
%% NatureName	-> \tanm{Poet Laureate} -> Title after name
%% Degrees	-> \dgr{MSc, PhD}
%% \author*[1,2]{\pfx{Dr} \fnm{Joergen W.} \spfx{van der} \sur{Ploeg} \sfx{IV} \tanm{Poet Laureate} 
%%                 \dgr{MSc, PhD}}\email{iauthor@gmail.com}
%%=============================================================%%

\author*[1]{\fnm{Gao Yu} \sur{Lee}}\email{GAOYU001@e.ntu.edu.sg}
\equalcont{These authors contributed equally to this work.}

\author[2]{\fnm{Tanmoy} \sur{Dam}}\email{tanmoy.dam@ntu.edu.sg}
\equalcont{These authors contributed equally to this work.}

\author[2,3]{\fnm{Md Meftahul} \sur{Ferdaus}}\email{mferdaus@uno.edu}
\equalcont{These authors contributed equally to this work.}

\author[1]{\fnm{Daniel Puiu} \sur{Poenar}}\email{EPDPuiu@ntu.edu.sg}

\author[2]{\fnm{Vu N.} \sur{Duong}}\email{vu.duong@ntu.edu.sg}

\affil*[1]{\orgdiv{School of Electrical and Electronic Engineering}, \orgname{Nanyang Technological University}, \orgaddress{\street{50 Nanyang Ave}, \postcode{639798}, \country{Singapore}}}

\affil[2]{\orgdiv{School of Mechanical and Aerospace Engineering}, \orgname{Nanyang Technological University}, \orgaddress{\street{65 Nanyang Drive}, \postcode{637460}, \country{Singapore}}}

\affil[3]{\orgdiv{Department of Computer Science}, \orgname{The University of New Orleans, New Orleans}, \orgaddress{\street{2000 Lakeshore Drive}, \postcode{LA 70148}, \country{USA}}}

% \affil[3]{\orgdiv{School of Mechanical and Aerospace Engineering}, \orgname{Nanyang Technological University}, \orgaddress{\street{Street}, \city{City}, \postcode{610101}, \state{State}, \country{Country}}}

%%==================================%%
%% sample for unstructured abstract %%
%%==================================%%

\abstract{Recent advancements have significantly improved the efficiency and effectiveness of deep learning methods for image-based remote sensing tasks. However, the requirement for large amounts of labeled data can limit the applicability of deep neural networks to existing remote sensing datasets. To overcome this challenge, few-shot learning has emerged as a valuable approach for enabling learning with limited data. While previous research has evaluated the effectiveness of few-shot learning methods on satellite-based datasets, little attention has been paid to exploring the applications of these methods to datasets obtained from Unmanned Aerial Vehicles (UAVs), which are increasingly used in remote sensing studies. In this review, we provide an up-to-date overview of both existing and newly proposed few-shot classification techniques, along with appropriate datasets that are used for both satellite-based and UAV-based data. Our systematic approach demonstrates that few-shot learning can effectively adapt to the broader and more diverse perspectives that UAV-based platforms can provide. We also evaluate some state-of-the-art few-shot approaches on a UAV disaster scene classification dataset, yielding promising results. We emphasize the importance of integrating explainable AI (XAI) techniques like attention maps and prototype analysis to increase the transparency, accountability, and trustworthiness of few-shot models for remote sensing. Key challenges and future research directions are identified, including tailored few-shot methods for UAVs, extending to unseen tasks like segmentation, and developing optimized XAI techniques suited for few-shot remote sensing problems. This review aims to provide researchers and practitioners with an improved understanding of few-shot learning’s capabilities and limitations in remote sensing, while highlighting open problems to guide future progress in efficient, reliable, and interpretable few-shot methods.}


%%\pacs[JEL Classification]{D8, H51}

%%\pacs[MSC Classification]{35A01, 65L10, 65L12, 65L20, 65L70}

\maketitle

\section{Introduction}\label{sec1}

\begin{figure}[hbt!]
    \centering
    \includegraphics[scale=0.75]{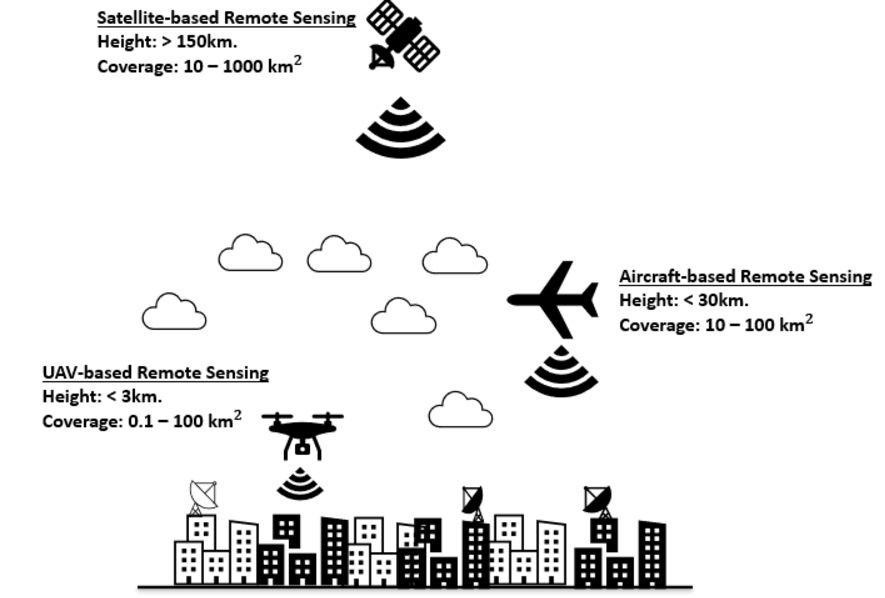}
    \caption{A simple pictorial illustration of the type of remote sensing platforms, as well as the typical height in which such platforms navigate. The values for their possible coverage area for an urban setting are also illustrated and are adapted from \cite{xiang2018mini}.}
    \label{fig:coverage_area}
\end{figure}

The last few decades saw significant advancements in remote sensing imaging technology. Remote sensing technologies nowadays encompass not only the traditional satellite-based platforms, but also include data collected from remote Unmanned Aerial Vehicles (UAVs). Figure \ref{fig:coverage_area} illustrates the typical height at which such platforms navigate as well as their estimated coverage area \cite{xiang2018mini} for an urban setting.  The modern airborne sensors that are attached to such platforms can cover and map a significant portion of the earth's surface with better spatial and temporal resolutions, making them essential for earth-based or environmental-based observations like geodesy and disaster relief. Automatic analysis of remote sensing images is usually multi-modal, meaning that optical, radar, or infrared sensors could be used, and such data could be distributed geographically and globally in an increasingly efficient manner. With advances in artificial intelligence, deep learning approaches have found their way into the remote sensing community, which, together with the increased in remote sensing data availability, has enabled more effective scene understanding, object identification, and tracking.

Convolutional Neural Networks (CNNs) have become popular in object recognition, detection, and semantic or instance segmentation of remote sensing images, typically using RGB images as input, which undergo convolution, normalization, and pooling operations. The convolution operation is effective in accounting for the local interactions between features of a pixel. While the remote sensing community has made great strides in multi-spectral satellite-based image classification, tracking, and semantic and instance segmentation, the limited receptive field of CNNs makes it difficult to model long-range dependencies in an image. Vision transformers (ViTs) was proposed to address this issue by leveraging the self-attention mechanism to capture global interactions between different parts of a sequence. ViTs have demonstrated high performance on benchmark datasets, competing with the best CNN-based methods. Consequently, the remote sensing community has rapidly proposed ViT-based methods for classifying high-resolution images. With pre-training weights and transfer learning techniques, CNNs and ViTs can retain their classification performance at a lower computational cost, which is essential for limited computational resources platforms such as UAVs.

However, both CNNs and ViTs required large training data samples for accurate classification, and some of these methods may not be feasible for critical tasks such as UAVs search-and-rescue. It would be beneficial, for instance, if the platforms were able to quickly identify and generalize disaster scene solely from analyzing a small subset of the captured frames. Few-shot classification approaches addressed the above needs, and in such approaches the goal is to enable the network to quickly generalize to unseen test classes in a more diverse manner given a small sets of training images. A framework like this closely resembles how the human brain learns in real life. Like ViTs, few-shot learning has also ignited new researches in remote sensing, and their applications to land cover classification in the RGB domain (\cite{deng2021cnns},\cite{zhang2021trs}) and hyperspectral classification (\cite{he2019hsi}, \cite{zhong2021spectral}) has been observed. The approaches have also been extended to object detection \cite{carion2020end} and segmentation \cite{xu2021efficient}. These emerging works are also as recent as that utilizing ViT. Since a review of ViT approaches for various domains in remote sensing \cite{aleissaee2022transformers} have been reported, a review of few-shot-based approaches in remote sensing is noteworthy to keep current interested researchers up-to-pace with the recent progresses in this area.

We have taken note that a related review has already been conducted in \cite{sun2021research}. A notable omission in the previous review is the failure to acknowledge the significance of interpretable machine learning models in this field. Integrating interpretable machine learning into remote sensing image classification can further enhance CNNs and ViTs' performance. By providing insights into the decision-making process of these models, interpretable machine learning can increase their transparency and accountability, which is particularly relevant in applications where high-stakes decisions are made based on their outputs, such as disaster response and environmental monitoring. For instance, saliency maps can be generated to highlight regions of images that are most relevant for the model's decision, providing visual explanations for its predictions. Furthermore, interpretable machine learning can aid in identifying potential biases and errors in the training data, as well as enhancing the robustness and generalization of the model. In remote sensing, interpretable machine learning can also facilitate the integration of expert knowledge into the model, enabling the inclusion of physical and environmental constraints in the classification process. This can enhance the accuracy and interpretability of the model, allowing for more informed decision-making. In short, the integration of interpretable machine learning in remote sensing image classification can provide a valuable tool for enhancing the transparency, accountability, and accuracy of CNNs and ViTs. By providing insights into the decision-making process of these models, interpretable machine learning can help build trust in their outputs and facilitate their use in critical applications.

The purpose of this additional review in this area is to address some more gaps that were not included in the previous review by \cite{sun2021research}. These gaps are as follows:

\begin{itemize}
\item Their focus in exploring remote sensing datasets was on satellite-based imagery and the few-shot learning techniques associated with such datasets. Nonetheless, with the emergence of UAV-based remote sensing datasets, we have detected a lack of consideration for the proposed works that have been applied and evaluated in such datasets. Furthermore, datasets and learning-based techniques associated with UAVs could also benefit from few-shot learning approaches due to their limited computational resources. This implies that data collected through UAVs would be constrained by a limited amount, thus emphasizing the need for efficient learning methods.
\item Quantitatively speaking, satellite-based remote sensing datasets offer a considerably wider field of view and greater coverage, allowing for the simultaneous capture of multiple object classes or labels in a single scene, an approach referred to as multi-label classification. On the other hand, the smaller coverage area of UAV-based remote sensing datasets often provides data that is suitable only for single-label image classification. Consequently, proposed methods that address such settings in the context of UAV-based remote sensing can be easily distinguished from those designed for multi-label classification, in contrast to works utilizing satellite-based remote sensing datasets. It is essential, therefore, to take into account the characteristics of the remote sensing dataset when devising and evaluating image classification methods in this field.
\item As has been emphasized and illustrated by \cite{sun2021research}, the utilization of few-shot learning-based techniques for remote sensing has been on the rise since 2012. As the aforementioned work was published in 2021, we can envisage that there will be an even greater proliferation of such approaches for remote sensing. In light of the dynamic nature of this research domain, our review aims to disseminate the most current and up-to-date information available on the topic. Through this approach, we seek to provide an improved understanding of the recent advances in few-shot learning-based methods for remote sensing, allowing for a comprehensive assessment of their potential applications and limitations. 
\end{itemize}

In summary, our main contributions in this review article are as follows: 
\begin{itemize}
  \item In this work, we present and holistically summarize the applications of few-shot learning-based approaches in both satellite-based and UAV-based remote sensing images, focusing on image classification alone, but extending the review work conducted by \cite{sun2021research} in terms of the explored remote sensing datasets. Our analysis serves to assist readers and researchers alike, allowing them to bridge gaps between current state-of-the-art image-based classification techniques in remote sensing, which may aid in promoting further progress in the field.
  
  \item As part of our discussion on the recent progress in the field of remote sensing regarding few-shot classification, we examined how CNNs and transformer-based approaches can be adapted to datasets, expanding the potential of these methods in this domain.
  
  \item Our work delved into a thorough discussion of the challenges and research directions concerning few-shot learning in remote sensing. We aimed to identify the feasibility and effectiveness of different learning approaches in this field, focusing on their potential applications in UAV-based classification datasets. Through this approach, we sought to shed light on the potential limitations and further research needed to improve the efficacy of few-shot learning-based techniques in the domain of remote sensing, paving the way for more advanced and sophisticated classification methods to be developed in the future.
  
  \item We also emphasized the significance of integrating XAI to improve transparency and reliability of few-shot learning-based techniques in remote sensing. Our objective was to offer researchers and practitioners a better comprehension of the possible applications and constraints of these techniques. We also aimed to identify novel research directions to devise more effective and interpretable few-shot learning-based methods for image classification in remote sensing.
\end{itemize}

\begin{figure}
    \centering
    \includegraphics[scale=0.2]{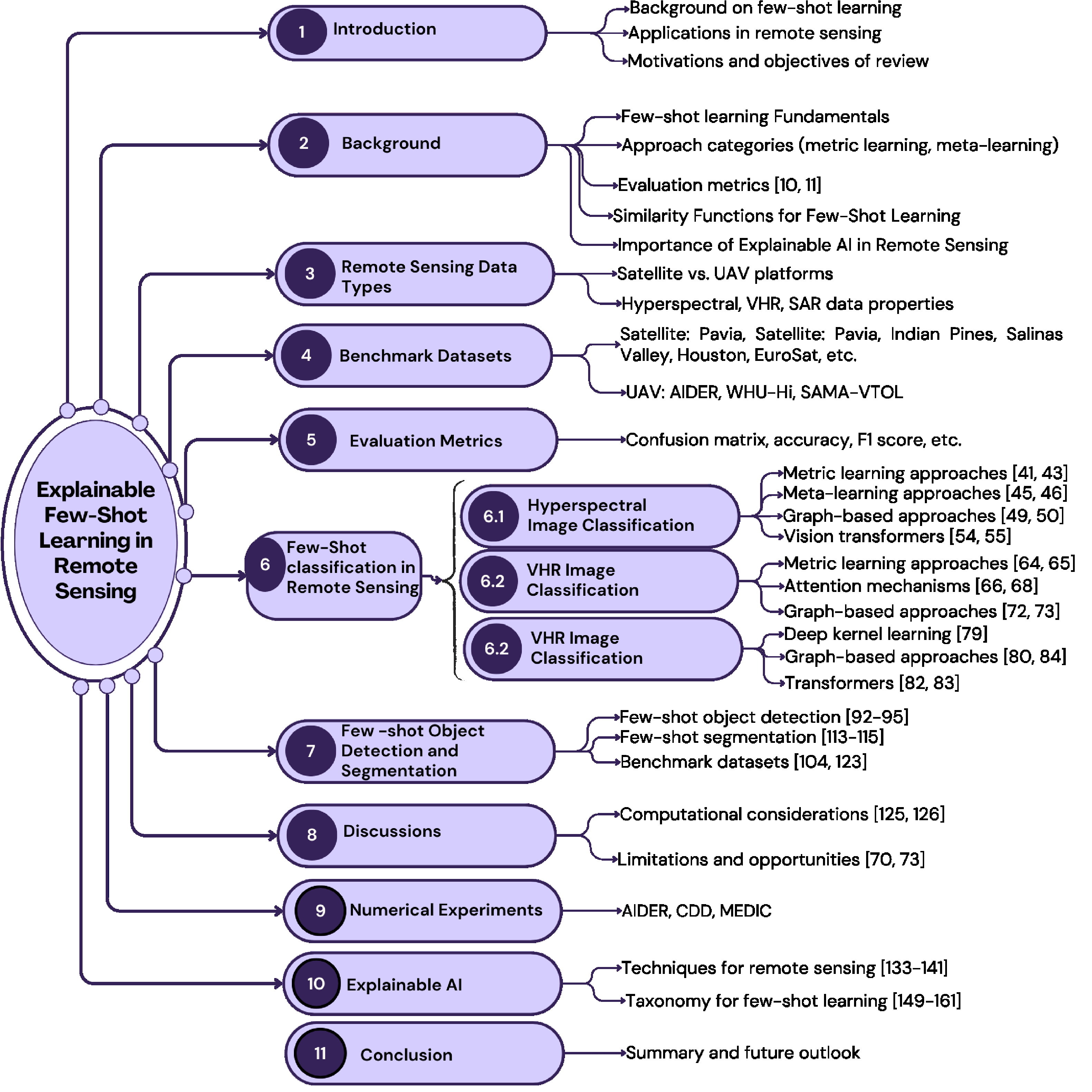}
    \caption{Overview of Explainable Few-Shot Learning in Remote Sensing. This illustration provides a high-level summary of the scope of our review on Explainable Few-Shot Learning techniques, applications, and challenges within Remote Sensing.}
    \label{fig:whole_taxo}
\end{figure}

The remainder of this paper is structured as follows: In Section 2, we provide a quick background on few-shot learning and present example networks. Section 3 discusses related review works in remote sensing, and Section 4, 5 and 6 provide brief highlights of the type of remote sensing data, common evaluation metrics utilized, and benchmark datasets commonly used, respectively. Section 7 delves into some up-to-date existing works on few-shot classification in the hyperspectral, Very High Resolution (VHR), and Synthetic Aperture Radar (SAR) data domain. In Section 8, we outline some implications and limitations of current approaches, and in Section 9, we quantitatively evaluate some existing methods on a UAV-based dataset, demonstrating the feasibility of such approaches for UAV applications. Finally, in Section 10, we conclude this review paper. An overview of the scope covered in our review of Explainable Few-Shot Learning for Remote Sensing is illustrated in Figure \ref{fig:whole_taxo}.

\section{Backgrounds}
Few-shot learning (FSL) is an emerging approach in the field of machine learning that allows models to acquire knowledge and make accurate predictions with limited training examples per class or context in a specific problem domain. In contrast to conventional machine learning techniques that demand vast quantities of training data, FSL aims to achieve comparable levels of performance using substantially fewer training examples. This ability to learn from scarce data makes FSL well-suited for applications where gathering sizable training sets may be prohibitively expensive or otherwise infeasible.

\begin{figure}[H]
\centering
\includegraphics[scale=0.17]{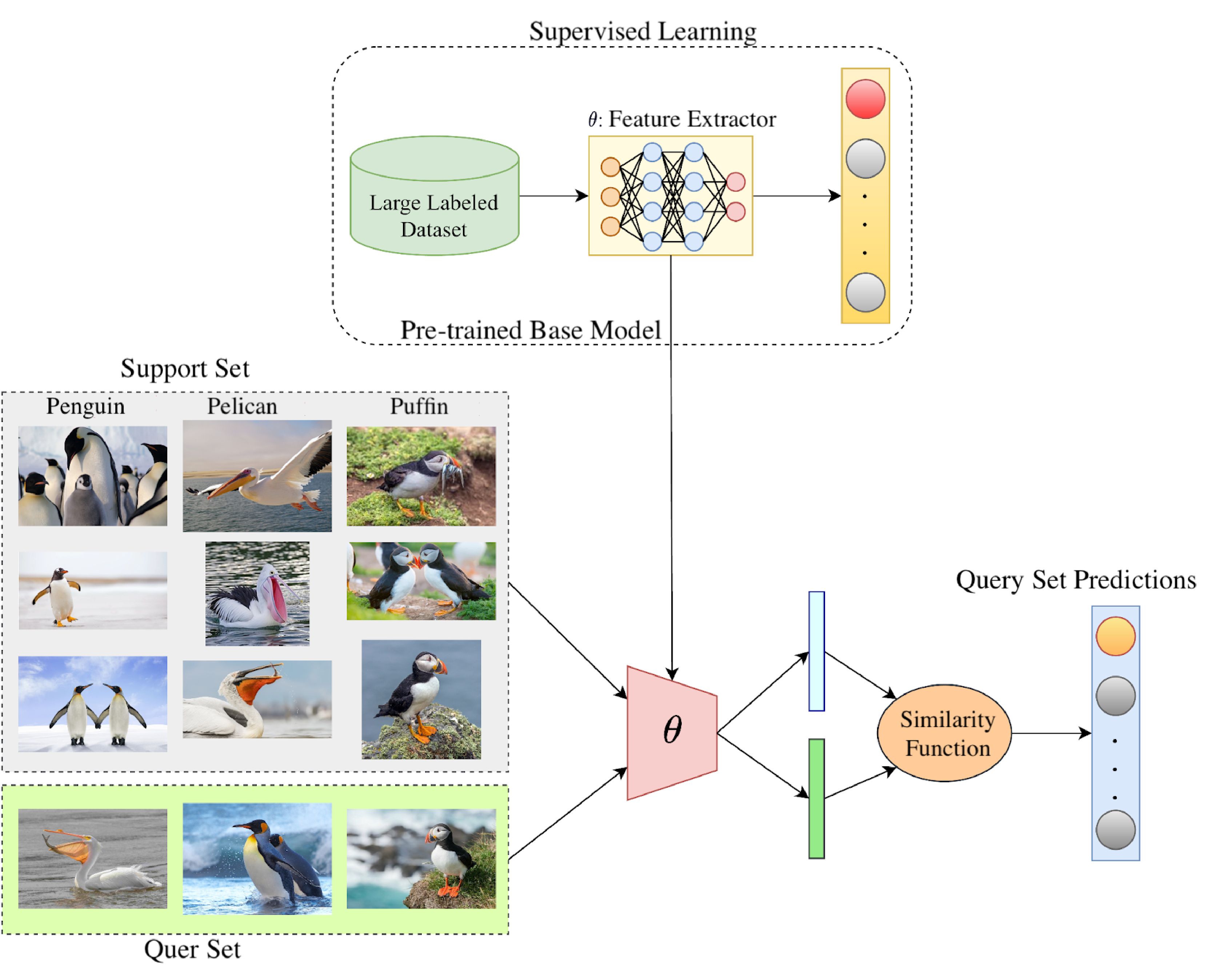}
\caption{Schematic of the few-shot learning approach, including pre-training a base model on ample data, constructing task episodes with support and query sets for new classes, fine-tuning the base model on the support set, and evaluating on the query set.}\label{fig:fsl_main}
\end{figure}

% In FSL, the input data are typically divided into two main categories: the support set and the query set. The support set consists of a small number of labeled examples that the model uses to learn the underlying patterns in the data, while the query set consists of a larger number of unlabeled examples that the model uses to evaluate its ability to generalize. The model is then trained to recognize the different classes or contexts in the query set, based on the patterns that it has learned from the support set.
In traditional machine learning, models are trained from scratch on large labeled datasets. In contrast, FSL aims to learn new concepts from just a few examples, leveraging transfer learning from models pre-trained on other tasks. First, a base model is pretrained on a large dataset for a task like image classification. This provides the model with general feature representations that can be transferred. Then for the new few-shot task, the pretrained model is used as a starting point. The support set of few labeled examples for the new classes is used to fine-tune the pretrained model. Typically only the last layer is retrained to adapt the model to the new classes, in order to leverage the pre-learned features. Finally, the adapted model is evaluated on the query set. The query set contains unlabeled examples that the model must make predictions for, based on the patterns learned from the small support set for each new class. This tests how well the model can generalize to new examples of the classes after adapting with only a few shots. To get a clearer view, this whole process is illustrated in Figure \ref{fig:fsl_main}. 

Approaches in FSL classification can often be categorized based on the number of novel categories needed for generalization, referred to as $N$, as well as the number of labeled samples or classes available in the support set for each of the $N$ novel classes, referred to as $k$. Generally, a lower value of $k$ makes it more challenging for the few-shot model to achieve high classification accuracy, as there is less supporting information in the support set to aid the model in making accurate predictions. This scheme is commonly referred to as `$N$-way $k$-shot learning scheme.' In instances where $k$ equals 1, such schemes are often referred to as one-shot learning. Additionally, in instances where $k$ equals 0, such schemes are often referred to as zero-shot learning.

Initial exploration of FSL in conjunction with unmanned aerial vehicle (UAV)-based thermal imagery was undertaken by \cite{liu2018real,masouleh2019development}. Their pioneering work demonstrated the potential of FSL for UAV-based tasks where limited onboard computational resources impose stringent constraints on model complexity and training data volume. The primary goal of FSL is to construct models that can identify latent patterns within a certain field using limited training examples, then utilize this learned knowledge to effectively categorize and classify new input. This capability closely mirrors human learning, where people can often understand the core of a new concept from just one or two examples. By reducing reliance on extensive training sets, FSL facilitates the development of machine learning systems applicable to data-scarce real-world problems across a broad range of domains.

% \textcolor{blue}{The first attempt to use deep learning with little data and UAV-based thermal images was carried out by \cite{liu2018real,masouleh2019development}.} Few-Shot Learning (FSL) is a subfield of machine learning that focuses on training models to effectively learn and generalize patterns from a few examples. In contrast to traditional machine learning approaches that require a vast amount of data for training, FSL aims to achieve the same level of accuracy using just a few training samples per context or class within the same domain. The goal of FSL is to enable models to effectively learn and generalize from a small set of training samples, while also being able to accurately recognize and classify new examples from the same domain.

% The challenge of FSL lies in its ability to generalize beyond the training samples, which requires models to capture the underlying patterns and relationships between the different classes or contexts. To accomplish this, FSL models are often designed with the ability to extract features that can capture essential information from input data, and can then use this information to make accurate predictions. 

\begin{figure}[H]
\centering
\includegraphics[scale=0.45]{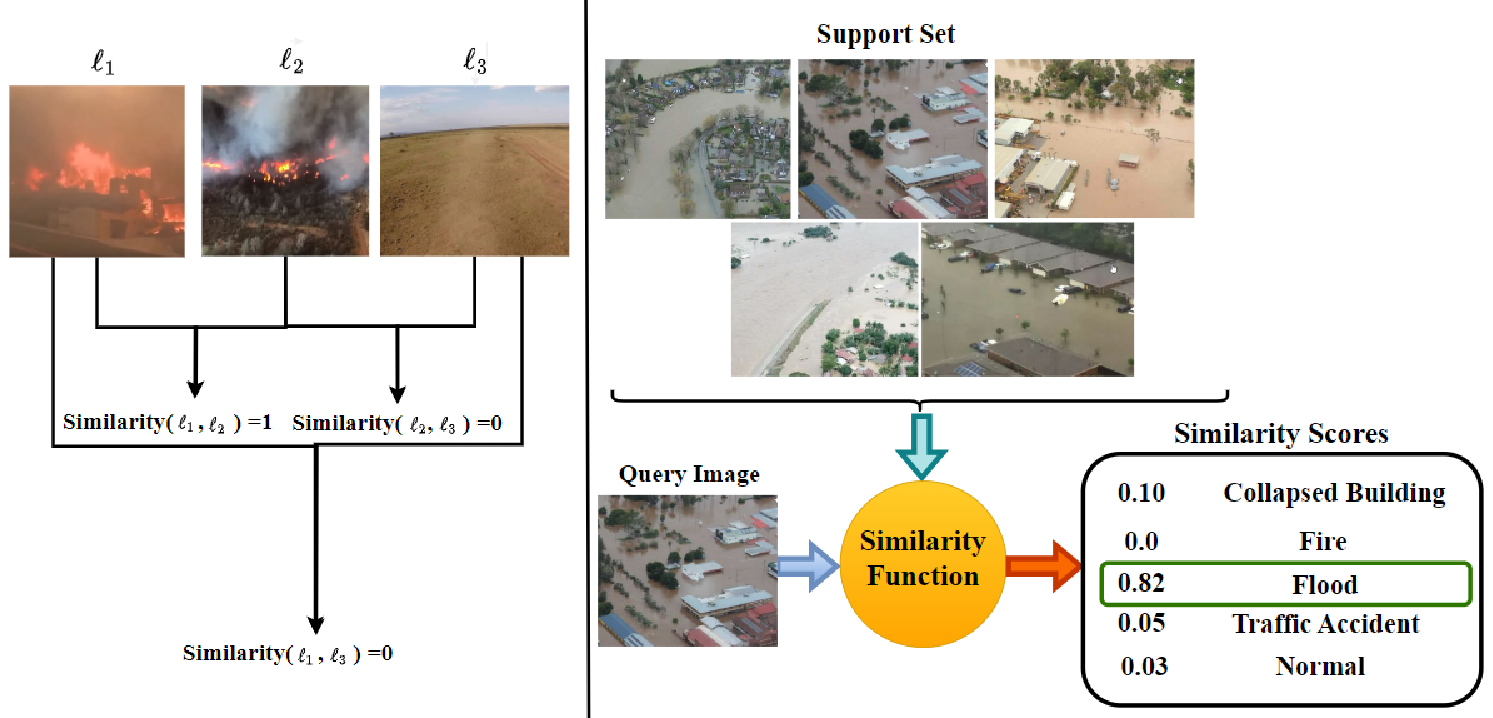}
\caption{{\textbf{(Left)} Similarity function as applied to each pair of images in the AIDER dataset \cite{kyrkou2020emergencynet}. The image on the left and middle constitute a fire disaster class, and the image on the right is a non-disaster class (normal). \textbf{(Right)} A query image of the flood disaster class is compared with the images from the support set via the similarity function and a correct class prediction is made based on the similarity score. In this case the flood disaster class is correctly predicted and classified.}\label{fig2}}
\end{figure}

\subsection{Similarity Functions for Few-Shot Learning}
A similarity function is a critical component of linking the support set and query set in few-shot learning. An example in the context of aerial disaster scene classification using the AIDER dataset \cite{kyrkou2020emergencynet} is illustrated in Figure \ref{fig2}. The left side of the figure shows how the similarity function evaluation is performed between each pair of images, with the left and middle images representing a fire disaster class and the right image representing a non-disaster class (or normal class). The right side of the figure shows how the similarity function can be used in conjunction with a query image and those from the support set to make a prediction on the correct class (flood) based on the similarity scores.

In few-shot learning, the choice of loss function is critical for enabling effective generalization from limited examples. Some commonly used loss functions include triplet loss, contrastive loss, and cross-entropy loss. The triplet loss helps models learn useful feature representations by minimizing the distance between a reference sample and a positive sample of the same class, while maximizing the distance to a negative sample from a different class. This allows fine-grained discrimination between classes. Contrastive loss is useful for training encoders to capture semantic similarity between augmented views of the same example. This improves robustness to input variations. Cross-entropy loss is commonly used for classifier training in few-shot models, enabling efficient learning from scarce labeled data. However, it can suffer from overfitting due to limited examples. Regularization methods such as label smoothing can help mitigate this. Other advanced losses like meta-learning losses based on model parameters have shown promise for fast adaptation in few-shot tasks. Overall, the choice of loss function plays a key role in addressing critical few-shot learning challenges like overfitting, feature representation learning, and fast generalization. Further research on specialized losses could continue improving few-shot performance.

For the scenario depicted on the left side of Figure \ref{fig2}, the triplet loss $L_{triplet}$ \cite{hoffer2015deep} is an example of a similarity function that could be used. The triplet loss involves comparing an anchor sample class to a positive sample class and a negative sample class. The goal is to minimize the Euclidean distance between the anchor and the positive class based on the similarity function $f$ and maximize the distance between the anchor and the negative class. This can be summarized mathematically in equation \ref{eq 1} as

\begin{equation}
L_{triplet} = \sum_{i}^{N} \norm{f(x_{i}^{a}) - f(x_{i}^{p})}_{2}^{2} -  \norm{f(x_{i}^{a}) - f(x_{i}^{n})}_{2}^{2} + \alpha,
\label{eq 1}
\end{equation}

In equation \ref{eq 1}, the anchor, positive, and negative class samples are denoted as $a$, $p$, and $n$, respectively. The index $i$ refers to the input sample index, $N$ denotes the total number of samples in the dataset, and $\alpha$ is a bias term acting as a threshold. The subscript 2 indicates that the evaluated Euclidean distance is the L2 loss, and the superscript 2 corresponds to squaring each parenthesis. The second term with a negative sign allows the maximization of the distance between the anchor and the negative class sample.

Networks that use the triplet loss for few-shot learning are also referred to as triplet networks. On the other hand, for comparing pairs of images, Siamese networks are commonly used. In such cases, the contrastive loss function $L_{contrastive}$ \cite{hadsell2006dimensionality} can be a better choice for defining similarity or loss, although the triplet loss can also be employed. The contrastive loss can be expressed mathematically as shown in equation \ref{eq2}:

\begin{equation}
\begin{split}
L_{contrastive}(W, y, (x_{1},x_{2})^{i} ) = \sum_{i}^{N} \frac{1}{2} (1-y)(D_W)^{2} + \frac{1}{2}(y)({max(0,m-D_W)}^{2}),
\end{split}
\label{eq2}
\end{equation}

In equation \ref{eq2}, $y$ denotes whether two data points, $x_{1}$ and $x_{2}$, within a given set $i$, are similar ($y$ = 0) or dissimilar ($y$ = 1). The margin term $m$ is user-defined, while $D_W$ is the similarity metric, which is given by:

\begin{equation}
D_W = \norm{f_{W}(X_{1}) - f_{W}(X_{2})}_{2}.
\label{eq3}
\end{equation}

Similarly to the previous method, the L2 loss-based Euclidean distance is used, where the first term in equation \ref{eq3} corresponds to similar data points and the second term corresponds to dissimilar ones. 

The third type of network for Few-Shot Learning can be realized as a prototypical network, as introduced by Snell et al. \cite{snell2017prototypical}. This method utilizes an embedding space in which samples from the same class are clustered together. In Figure 3, an example is provided to demonstrate this concept. For each cluster, a typical class prototype is computed as the mean of the data points in that group. The calculation of the class prototype can be expressed mathematically as shown in equation \ref{eq4}:

\begin{equation}
v^{(k)} = \frac{1}{N_{s}}\sum_{i}^{N_s} f_{\phi}(x_{i}^{k}),
\label{eq4}
\end{equation}

The equation \ref{eq4} represents the prototypical network, a third type of network for FSL. The class prototype, computed as the mean of the data points belonging to the same group in the embedded space, is denoted by $v^{(k)}$, where $k$ represents the class. The set of support images for class $k$ is represented by $x_{i}^{k}$, and the embedding function by $f_{\phi}$, which is different from the similarity function $f$ described earlier.

\begin{figure}[hbt!]
    \centering
    \includegraphics[width=11.0cm, height = 4.5cm]{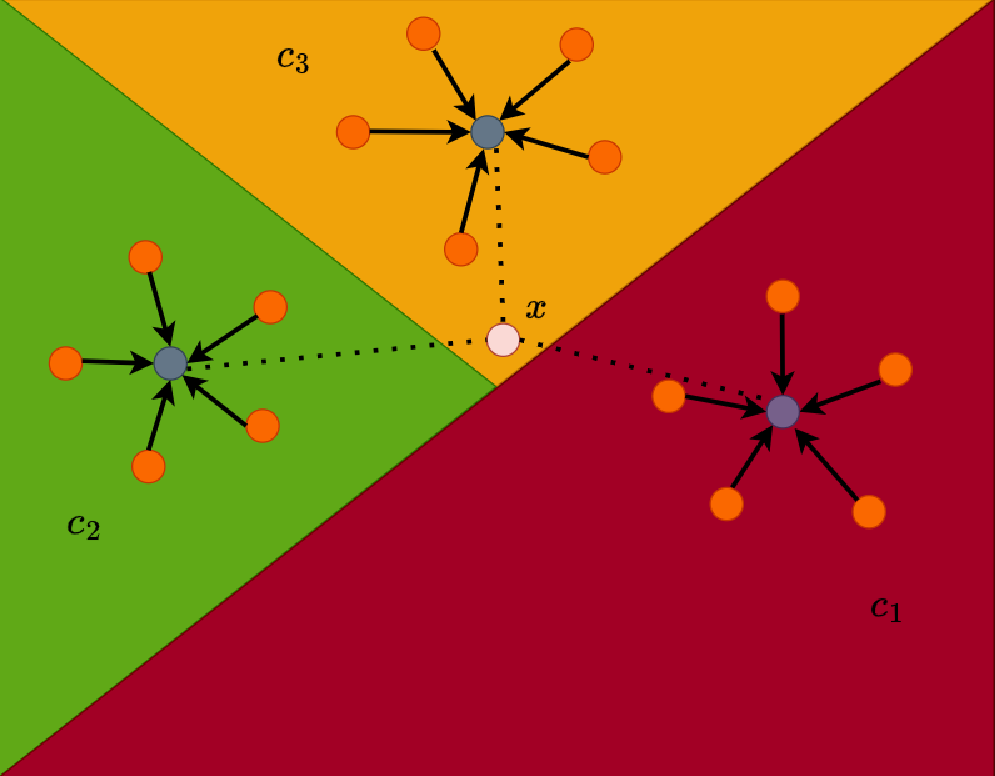}
    \caption{The embedding space in a prototypical network comprising of 3 classes $c_{1}$, $c_{2}$ and $c_{3}$ (denoted as green, dark red and orange respectively). The mean of $k$ samples (5 in the diagram) for each class serves as the center of the cluster, and a given data point $x$ is evaluated via the euclidean distance to determine which set of cluster it belongs based on the minimum distance criteria.}
    \label{fig:clusters}
\end{figure}

The prototypical network and Siamese or triplet networks are different few-shot learning approaches that compare query and support samples in different ways. While Siamese or triplet networks directly compare query and support samples in pairs or triplets, the prototypical network compares the query samples with the mean of their support set. This is achieved by calculating the prototype representation of each class in the embedded metric space, which is the average of the feature vectors of all the support samples for that class. This can be visualized in Figure \ref{fig:clusters}. However, for one-shot learning, where only a single support sample is available for each class, the three approaches become equivalent, as the prototype representation becomes identical to the support sample representation. Overall, the choice of few-shot learning approach may depend on the dataset's specific characteristics and the available support samples.

\subsection{Importance of explainable AI in remote sensing}
Remote sensing and analysis of satellite imagery has progressed rapidly thanks to artificial intelligence and machine learning. Machine learning models can identify objects and patterns in huge amounts of satellite data with incredible accuracy, surpassing human capabilities. However, these complex machine learning models are often considered "black boxes" - they provide highly accurate predictions and detections but it is unclear why they make those predictions.

Explainable AI is an emerging field of study focused on making machine learning models and their predictions more transparent and understandable to humans. Explainable AI techniques are essential for applications like remote sensing where decisions could have serious real-world consequences \cite{kakogeorgiou2021evaluating}. For example, a machine learning model that detects signs of natural disasters like wildfires in satellite images needs to provide an explanation for its predictions so that human operators can verify the findings before taking action. There are several approaches to making machine learning models used for remote sensing more explainable. 

\begin{itemize}
    \item \textbf{Highlighting important features:} Techniques like saliency maps can highlight the most important parts of an image for a machine learning model’s prediction. For example, computer vision models could highlight the features they use to detect objects in satellite images, allowing correction of errors. Similarly, anomaly detection models could point to regions that led them to flag unusual activity, enabling verification of true positives versus false alarms.
    \item \textbf{Simplifying complex models:} Complex machine learning models can be converted into simplified explanations that humans can understand, like logical rules and decision trees. For instance, deep reinforcement learning policies for navigating satellites could be expressed as a simplified set of if-then rules, revealing any flawed assumptions. These simplified explanations make the sophisticated capabilities of machine learning more accessible to domain experts.
    \item \textbf{Varying inputs to understand responses:} Another explainable AI technique is to systematically vary inputs to a machine learning model and observe how its outputs change in response. For example, generative models that create new realistic satellite images could be evaluated by generating images with different attributes to determine their capabilities and limitations. Analyzing how a model’s predictions vary based on changes to its inputs provides insights into how it works and when it may produce unreliable results.
\end{itemize}

Overall, explainable AI has the potential to build trust in machine learning systems and empower humans to make the best use of AI for applications like remote sensing.
Making machine learning models explainable also allows domain experts to provide feedback that can improve the models. For example, experts in remote sensing may notice biases or errors in a machine learning model’s explanations that could lead the model astray. By providing this feedback, the experts can help data scientists refine and retrain the machine learning model to avoid those issues going forward.

In short, explainable AI has significant promise for enabling machine learning and remote sensing to work together effectively. By making machine learning models and predictions transparent, explainable AI allows:
\begin{itemize}
    \item Humans to verify and trust the outputs of machine learning models before taking consequential actions based on them.
    \item Domain experts to provide feedback that improves machine learning models and avoids potential issues.
    \item A better understanding of the strengths, weaknesses and limitations of machine learning that can guide how the technology is developed and applied in remote sensing.
\end{itemize}
Explainable AI will be key to ensuring machine learning is used responsibly and to its full potential for remote sensing and beyond. Building partnerships between humans and AI can lead to a future with technology that enhances human capabilities rather than replacing them.

\section{Type of remote sensing sensor data}
Remote sensing data is typically acquired from satellite or unmanned aerial vehicle (UAV) platforms, and the characteristics of the data can vary greatly depending on the specific platform and sensor used. They can be classified according to their spatial, spectral, radiometric, and temporal resolutions, as discussed in \cite{aleissaee2022transformers} and \cite{sun2021research}.

\begin{itemize}
\item \textbf{Spatial resolution}: The spatial resolution of remote sensing data is often limited by the size and altitude of the sensor platform, as well as the resolution of the sensor itself. For example, satellite-based sensors typically have a lower spatial resolution than UAV-based sensors, due to their higher altitude and larger coverage area.

\item \textbf{Spectral resolution}: Spectral resolution refers to the range of wavelengths that a remote sensing sensor can detect, as well as the sampling rate at which it collects data across this range. Different sensors have different spectral characteristics, and the spectral resolution of a sensor can have a significant impact on its ability to distinguish different features or objects in the scene.

\item \textbf{Radiometric resolution}: Radiometric resolution is related to the sensitivity of the sensor and the number of bits utilized for signal representation. A higher radiometric resolution means that the sensor is able to capture a wider range of signal strengths and more accurately represent the scene being imaged.

\item \textbf{Temporal resolution}: Temporal resolution is a critical characteristic of remote sensing data, as it can enable the tracking of changes in a scene over time. The frequency with which images are collected, as well as the length of time over which they are collected, can impact the ability of remote sensing systems to detect and monitor changes in the scene, such as vegetation growth or land use changes. 
\end{itemize}

Understanding the various characteristics of remote sensing data is important for developing effective machine learning approaches, as different methods may be better suited to different types of data. For example, models that perform well on high-resolution satellite imagery may not perform as well on lower-resolution UAV data, and vice versa. By considering the characteristics of the data and tailoring machine learning approaches to the specific problem at hand, researchers can develop more accurate and effective models for remote sensing applications.

\begin{figure}[hbt!]
    \centering
    \includegraphics[width=11.0cm, height = 4.5cm]{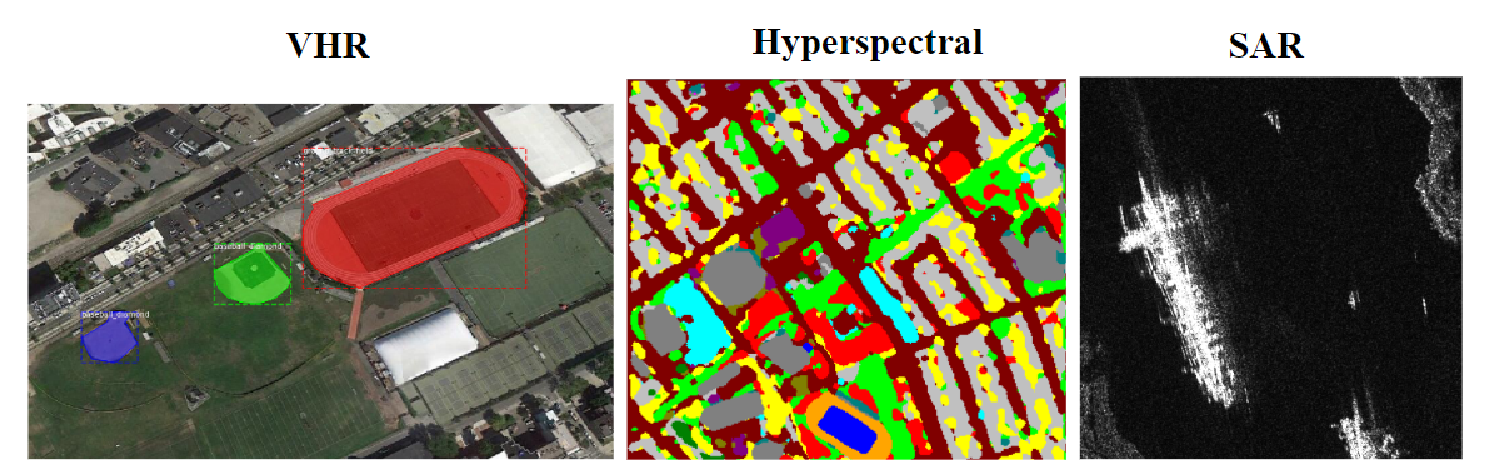}
    \caption{Illustration of each remote sensing image category. The VHR image is from the VHR-10 dataset \cite{su2019object}, the hyperspectral image is from the Houston dataset \cite{contest2013ieee}, and the SAR image is from the HR-SID dataset \cite{wei2020hrsid}.}
\end{figure}

A classification of image data types can also be made based on three categories, namely Very High-Resolution Imagery, Hyperspectral Imagery, and Synthetic Aperture Radar Imagery, as discussed in \cite{aleissaee2022transformers} and \cite{sun2021research}. Figure 4 provides some examples of such imagery.

\begin{itemize}
\item \textbf{Very High Resolution (VHR) imagery}: Very High Resolution (VHR) imagery is often captured through the use of VHR satellite sensors, which are designed to capture images with an extremely high level of detail. This level of detail can be particularly beneficial for a number of different applications, including object detection and tracking, as well as emergency response operations. As the technology behind optical sensors continues to advance, the spatial resolution obtained from these sensors becomes even finer, allowing for even greater levels of detail to be captured in these images. This, in turn, can lead to even more accurate object detection and tracking, as well as more effective emergency response operations that are better able to respond to events as they unfold in real time. 

\item \textbf{Hyperspectral imagery}: In addition to the optical electromagnetic spectrum that is often represented by the RGB color channels, remote sensing signals and imagery can also be obtained and analyzed in other parts of the spectrum, including the infrared (IR) and ultraviolet (UV) regions. In particular, the IR spectrum can be further categorized into near, mid or far-infrared, and the corresponding images captured in these ranges are known as hyperspectral imagery. This type of imagery goes beyond the three color channels of optical images and contains more spectral information, enabling the unraveling of the composition of the object of interest, both physically and chemically. As such, hyperspectral images are particularly useful for environmental and earth-science-based research, as they can provide detailed information on factors such as vegetation health, mineral composition, and water quality. By analyzing this spectral information, researchers can gain a deeper understanding of the earth's surface, as well as monitor changes and anomalies that may indicate potential issues.

\item \textbf{Synthetic Aperture Radar (SAR) imagery}: By utilizing the process of emission and reception of electromagnetic waves on the Earth, radar-based remote sensing can be accomplished. Such remote sensing techniques can acquire high spatial resolution images regardless of weather conditions, and are widely applicable in numerous domains. In particular, Synthetic Aperture Radar (SAR) based images have been utilized in diverse fields, including disaster management, hydrology and forestry, due to their ability to provide high-quality images regardless of atmospheric conditions, time of day or season. SAR-based imagery can thus be a valuable source of information for remote sensing-based research and applications. 
\end{itemize}

\section{Benchmark remote sensing datasets for evaluating learning models}
In this section, we will provide a brief overview of commonly used benchmark datasets that evaluate algorithms in remote sensing. The datasets are categorized and listed based on the type of remote sensing data and platforms they were collected from. It is important to note that these datasets are frequently used by researchers to evaluate and benchmark their algorithms, and although not included in the survey works by \cite{aleissaee2022transformers}, they are essential for this review.

\subsection{Hyperspectral image dataset}

\subsubsection{\textbf{Satellite-Based Data}}

Most of the datasets described here are more tailored for multi-label image classification, although a few single label-based classification dataset exist.

\begin{itemize}
\item \textbf{Pavia} \cite{GIC,dam2022developing}: The Pavia University research team created a hyperspectral image dataset with images consisting of 610 $\times$ 610 pixels and 103 spectral bands. Each image in the dataset is a classification map with 9 classes that include mostly urban contexts such as bitumen, brick, and asphalt. The dataset comprises 42,776 labeled images and is specifically designed for multi-label classification.
\item \textbf{Indian Pines} \cite{GIC,dam2020mixture}: The dataset contains hyperspectral images of a particular landscape in Indiana. It is a multi-label classification dataset where each map consists of 145 $\times$ 145 pixels and 224 spectral bands. There are 16 semantic labels available for each map, and the dataset has a total of 10,249 samples.
\item \textbf{Salinas Valley} \cite{GIC}: The Salinas Valley dataset consists of hyperspectral images collected from California, with multi-label classification maps of pixel size 512 $\times$ 217 and 224 spectral bands, similar to the Indian Pines dataset. There are 16 semantic classes with 54,129 samples. A subset of the Salinas dataset, referred to as \textit{Salinas-A}, includes only 86 $\times$ 86 image pixels of 6 classes, with a total of 5,348 samples.
\item \textbf{Houston} \cite{contest2013ieee}: The Hyperspectral Image Analysis group in collaboration with the NSF Funded Center for Airborne Laser Mapping (NCALM) has acquired images across the University of Houston. This dataset comprises 16 semantic classes of urban objects such as highways, railways, and tennis courts, unlike the Botswana, Indian Pines, and Salinas Valley datasets. The images have 144 spectral bands in the 380 nm to 1050 nm region, and each image has a pixel size of 349 $\times$ 1905. The dataset is designed for evaluating multi-label image classification.
\item \textbf{BigEarthNet} \cite{sumbul2019bigearthnet}: The dataset consists of pairs of Sentinel-2 images captured by a multi-spectral sensor, with 590326 pairs collected from 10 European countries. Each image in the pair has a size of 120 $\times$ 120 pixels and covers 13 spectral bands. The dataset is annotated with multiple land-cover classes or labels, making it suitable for multi-label classification evaluation.
\item \textbf{EuroSat} \cite{helbersat}: The dataset consists of images obtained from the Sentinel-2 satellite, covering 13 spectral bands with 10 classes and 27,000 labeled samples. It is utilized for evaluating single-label-based land cover and land use classification. Each image has a pixel size of 64 $\times$ 64.
\item \textbf{SEN12MS} \cite{schmitt2021remote}: The dataset comprises 180,662 images captured from Sentinel-1 and Sentinel-2, with four cover types categorized using different classification schemes. Each image is of size 256 $\times$ 256 and contains different spectral bands. The images are annotated by multiple land-cover labels, but the primary objective is to use these labels to infer the overall context of the scene, such as forest, grasslands, or savanna, making it suitable for single label-based scene classification. It is important to note that Sentinel-1 images are SAR images, making the dataset useful for SAR-based map classification as well.\\
\end{itemize} 

\subsubsection{\textbf{UAV-based dataset}}

\begin{itemize}
\item \textbf{WHU-Hi} \cite{hu2020whu}: The WHU-Hi dataset, which stands for Wuhan UAV-borne Hyperspectral Image, consists of UAV-based images of various crop types gathered in farming areas in Hubei province, China. It is divided into three sub-datasets: WHU-Hi-LongKou, WHU-Hi-HanChuan, and WHU-Hi-Honghu, each with different individual image sizes, numbers of labels/classes, and spectral bands, which are explained in Table 2. The dataset is suitable for evaluating multi-label classification algorithms.
\end{itemize}

\subsection{VHR image-based dataset}

\subsubsection{\textbf{Satellite-based datasets}}

\begin{itemize}
\item \textbf{UC Merced Landuse} \cite{yang2010bag}: The dataset was designed for single-label land use classification and comprises 2100 RGB images, each of size 256 $\times$ 256 pixels. The dataset consists of 21 classes, predominantly related to urban land use.
\item \textbf{ISPRS Potsdam} \cite{gerke2014use}: The International Society of Photogrammetry and Remote Sensing (ISPRS) developed a dataset for algorithmic evaluation of multi-label map classification. The dataset comprises 38 patches/images. The pixel size of each patch is 6000 $\times$ 6000.
\item \textbf{ISPRS Vaihingen} \cite{gerke2014use}: The dataset was created for multi-label map classification and includes 33 patches/images of varying sizes. The pixel size of each patch is 2494 $\times$ 2064.
\item \textbf{RESISC45} \cite{cheng2017remote}: The Northwestern Polytechnical University (NWPU) created a dataset for single-label image scene classification. The dataset contains 31,500 images categorized into 45 classes, with each class consisting of 700 images. The pixel size of each image is 256 $\times$ 256. 
\item \textbf{WHU-RS19} \cite{xia2010structural}: The dataset is created using satellite images obtained from Google Earth and contains 19 semantic classes, with approximately 50 samples per class. Samples from same class are extracted from different regions with varying resolutions, scales, orientations, and illuminations. Each image in the dataset is 600 x 600 pixels in size. It is intended for the purpose of single label-based image scene classification.
\item \textbf{AID} \cite{xia2017aid}: The Aerial Image Database (AID) is a collection of 10,000 satellite images gathered from Google Earth, each sized 600 $\times$ 600 pixels. The dataset includes 30 classes primarily related to urban environments. As with the RESISC45 and WHU-RS19 datasets, AID is used for single-label image scene classification purposes..\\
\end{itemize}

\subsubsection{\textbf{UAV-based dataset}}

\begin{itemize}
\item \textbf{AIDER} \cite{kyrkou2020emergencynet}: The Aerial Image Database for Emergency Response (AIDER) is a collection of 8540 UAV images categorized into four disaster categories - collapsed buildings, fire, flood, and traffic accidents, along with a non-disaster category labeled as "normal" \cite{lee2023watt}. This is one of the first UAV-based datasets that can be used as a benchmark for visual-based humanitarian aid or search-and-rescue operations in the RGB spectrum.
\item \textbf{SAMA-VTOL} \cite{bayanlou2021multi}: The SAMA-VTOL aerial image dataset is a new dataset developed from images captured by UAVs. This dataset was created to support a broad spectrum of scientific projects within the field of remote sensing. It is particularly useful for research projects focused on 3D object modeling, urban and rural mapping, and the processing of digital elevation and surface models. The objective is to provide high-resolution, low-cost data that contribute to a better understanding of both urban and rural scenes for various applications.
\end{itemize}

\subsection{SAR image-based dataset}
% \subsubsection{\textbf{Satellite-Based Data}}
\begin{itemize}
\item \textbf{MSTAR} \cite{wang2015application}: This dataset consists of 5950 X-band spectral images, each with a size of 128 $\times$ 128 pixels, and categorized into 10 classes. It is designed specifically for military object recognition and classification. 
\item \textbf{OpenSARShip} \cite{huang2017opensarship}: The dataset includes 11,346 chips of ships captured by C-band SENTINEL-1 SAR imagery, belonging to 17 ship types, and collected from 41 images. Each chip is labeled with automatic identification system messages indicating different environmental conditions. The image sizes of the chips range from 30 $\times$ 30 to 120 $\times$ 120 pixels. \\
\end{itemize}

It is evident that there are fewer SAR-based benchmark datasets compared to hyperspectral or VHR-based image datasets. According to Fu et al. \cite{fu2021few}, collecting SAR-based images with fine annotation is more challenging due to the difficulty of acquisition and the tedious and time-consuming process of interpreting and labeling such images. Furthermore, Rostami et al. \cite{rostami2019sar} stated that the devices used for generating SAR images are costly, and the data accessibility is strictly regulated due to its classification.

In Table 2, we have summarized the discussion on the available datasets, highlighting the data type, number of images and classes, pixel sizes, spectral bands (if any), platform, and classification method.

\begin{table}
\caption{Summary of some datasets commonly utilized for few-shot learning algorithmic evaluation in the domain of remote sensing.\label{tab2}}
  \resizebox{\textwidth}{!}{%
		\begin{tabular}{p{3cm}p{3cm}p{3cm}p{2cm}p{2cm}p{2cm}p{2cm}p{3cm}}  % Change this.
		\hline
			\textbf{Datasets} & \textbf{Data Type} & \textbf{Platforms} & \textbf{Images/ Samples} & \textbf{No. of classes} & \textbf{Image Size} & \textbf{Spectral Bands} & \textbf{Classification Type}\\ \hline
		
            Pavia & Hyperspectral & Satellite & 42,776 & 9 & 610 $\times$ 610 & 103 & Multi-label \tabularnewline
            Indian Pines & Hyperspectral & Satellite & 10,249 & 16 & 145 $\times$ 145 & 224 & Multi-label \tabularnewline
            Salinas Valley & Hyperspectral & Satellite & 54,129 & 16 & 512$\times$ 127 & 224 & Multi-label \tabularnewline
            Houston & Hyperspectral & Satellite & 504,712 & 16 & 349 $\times$ 1905 & 144 & Multi-label \tabularnewline
            BigEarthNet & Hyperspectral & Satellite & 590,326 & 12 & 120 $\times$ 120 & 13 & Multi-label \tabularnewline
            EuroSat & Hyperspectral & Satellite & 27,000 & 10 & 64 $\times$ 64 & 13 & Single-label \tabularnewline
            SEN12MS  & Hyperspectral & Satellite & 180,662 & 10 & 256 $\times$ 256 & 15 & Single-label \tabularnewline
            WHU-Hi-LongKou & Hyperspectral & UAV & 204,542 & 9 & 550 $\times$ 400 & 270 & Multi-label \tabularnewline
            WHU-Hi-HanChuan & Hyperspectral & UAV & 144,788 & 16 & 1217 $\times$ 303 & 274 & Multi-label \tabularnewline
            WHU-Hi-HongHu & Hyperspectral & UAV & 70,874 & 22 & 940 $\times$ 475 & 270 & Multi-label \tabularnewline
      
            UC Merced Landuse & VHR & Satellite & 2,100 & 21 & 256 $\times$ 256 & RGB & Single-label \tabularnewline
            ISPRS Potsdam & VHR & Satellite & 38 & 6 & 6,000 $\times$ 6000 & RGB & Multi-label \tabularnewline
            ISPRS Vaihingen & VHR & Satellite & 33 & 6 & 2,494 $\times$ 2064 & RGB & Multi-label \tabularnewline
            RESISC45 & VHR & Satellite & 31,500 & 45 & 256 $\times$ 256 & RGB & Single-label \tabularnewline
            WHU-RS19 & VHR & Satellite & 1,005 & 19 & 600 $\times$ 600 & RGB & Single-label \tabularnewline
            AID & VHR & Satellite & 10,000 & 30 & 600 $\times$ 600 & RGB & Single-label \tabularnewline
            AIDER & VHR & UAV & 8,540 & 5 & Varies & RGB & Single-label \tabularnewline
            \textcolor{blue}{SAMA-VOL} & Orthophoto & UAV & 120 & 8-10 & 2000 $\times$ 2000 & - & Multi-label \tabularnewline       
            MSTAR & SAR & Satellite & 5,950 & 10 & 128 $\times$ 128 & X-band & Single-label \tabularnewline
            OpenSARShip & SAR & Satellite & 11,346 & 17 & varies & C-band & Single-label \tabularnewline
            \hline
		\end{tabular}} 

\end{table}

\section{Evaluation metrics for few-shot remote sensing task}
Before delving into the various approaches in a few-shot remote sensing task, we highlight in this section some evaluation metrics that are more suited for few-shot learning task. The data distribution would display some degree of imbalance between the training set and the test set for small-sample size unlike typical learning-based tasks, and hence appropriate metrics addressing such imbalance would need to be invoked. We illustrate in Table 1 the various metrics along with a brief overview. The metrics are the confusion matrix, precision, recall, F1 score, Overall Accuracy (OA), Average Accuracy (AA), Kappa coefficient $\kappa$, and PR curve. (5)-(9) mathematically describe some of the metrics as indicated in the respective equations.

\begin{equation} \label{eq5}
\begin{split}
Precision = \frac{TP}{TP+FP}, \quad Recall = \frac{TP}{TP+FN},
\end{split}
\end{equation} 

\begin{equation} \label{eq6}
\begin{split}
\bar{F1}= \frac{2}{N_{classes}}\sum_{i=1}^{N_{classes}}\frac{Precision*Recall}{Precision + Recall},
\end{split}
\end{equation} 

\begin{equation} \label{eq7}
\begin{split}
OA = \frac{TP+TN}{TP+TN+FP+FN},
\end{split}
\end{equation} 

\begin{equation} \label{eq8}
\begin{split}
AA = \sum_{i=1}^{N}\frac{TP_{i}+TN_{i}}{TP_{i}+FN_{i}+TN_{i}+FN_{i}},
\end{split}
\end{equation} 

\begin{equation} \label{eq9}
\begin{split}
\kappa = \frac{2\times((TP \times TN)-(FN\times FP))}{(TP+FP) \times (FP+TN) \times (TP+FN) \times (FN+TN)}.
\end{split}
\end{equation} 

\begin{table}
\caption{Evaluation metrics commonly utilized in few-shot learning-based approaches.\label{tab1}}
 \resizebox{\textwidth}{!}{%
\begin{tabular}{p{3cm}p{6cm}}
\hline
\textbf{Metrics}& \textbf{Overview}	\\
\hline
Confusion Matrix & Summary of correct and incorrect predictions by a classifier in matrix form.\\
Precision & Ratio of true positive and the total predicted positives.\\
Recall & Ratio of true positive and the total ground-truth positives. \\ 
F1 Score & Weighted average of precision and recall.\\
Overall Accuracy (OA) & Sum of true positives plus true negatives divided by the total number of class sample involved. \\
Average Accuracy (AA) & Average computed accuracy per-class.\\

Kappa Coefficient ($\kappa$) & Measures of statistical agreement between existing and random classification results.\\

PR Curve & Statistical graph (y-axis: precision, x-axis: recall) \\
\hline
\end{tabular}}
\end{table}

The variables $TP$, $FP$, $TN$, and $FN$ in the previous equations represent true positive, false positive, true negative, and false negative classes, respectively. In equation (6), $N_{classes}$ refers to the total number of classes that are taken into consideration.

\section{Recent few-shot learning techniques in remote sensing}

In the domain of remote sensing, the intersection with computer vision has received considerable attention and research interest, as evinced by numerous works such as those undertaken by \cite{aleissaee2022transformers} and \cite{tuia2011survey}, which delve into and assess diverse active machine learning frameworks. Moreover, the intricacies of hyperspectral image classification and contemporary developments in machine learning and computer vision methods are explored by \cite{camps2013advances}. A comprehensive and exhaustive analysis of deep learning algorithms utilized for processing remote sensing images, while detailing current practices and available resources, is provided by \cite{zhu2017deep}. Furthermore, \cite{aleissaee2022transformers} presents an overview of Vision Transformer-based approaches to remote sensing, with a specific focus on very high-resolution, hyperspectral, and radar imaging. In this review, our specific focus lies on recent breakthroughs in the realm of few-shot learning techniques for remote sensing imaging. We seek to provide an in-depth exploration of the implications of such advancements for scene classification and comprehension in both satellite-based and UAV-based data collection platforms. The incorporation of explainable AI can aid in understanding the reasoning behind classification results, providing more transparency and confidence in decision-making processes.

\subsection{Few-Shot learning in hyperspectral images classification}
In the field of remote sensing, few-shot learning has gained significant traction, as highlighted in the introductory section. In this particular section, we shall concentrate on the techniques put forward for both single-label and multi-label remote sensing classification within the context of both satellite and UAV-based platforms. It is worth noting that, unless indicated otherwise, all the evaluation metrics employed in the studies under review in this section encompass OA, AA, and $\kappa$.

The MDL4OW model, presented by \cite{liu2020few}, employs a few-shot based deep learning architecture to classify five unknown classes through training on nine known classes. Notably, the proposed model departs from traditional centroid-based methods, instead utilizing extreme value theory from a statistical model, as depicted in Figure 3. Furthermore, the authors introduced a novel evaluation metric, the mapping error, which is particularly sensitive to imbalanced classification scenarios frequently encountered in hyperspectral remote sensing image datasets. Mathematical expression of the mapping error for $C$ classes is provided in \ref{eq10}, subject to constraints expressed in \ref{eq11} and \ref{eq12}.

\begin{equation} \label{eq10}
\begin{split}
Error = \frac{\sum_{i=1}^{C} \norm{A_{p,i} - A_{gt,i}}}{\sum_{i=1}^{C} A_{gt,i}},
\end{split}
\end{equation} 

\begin{equation} \label{eq11}
\begin{split}
A_{i} \geq 0,
\end{split}
\end{equation} 

\begin{equation} \label{eq12}
\begin{split}
\sum_{i=1}^{C+1} A_{p,i} = \sum_{i=1}^{C+1} A_{gt,i},
\end{split}
\end{equation} 

The mathematical expression of \ref{eq10} represents the mapping error, where $A_{p,i}$ signifies the predicted area of the $i$th class and $A_{gt,i}$ denotes the corresponding ground-truth area. Here, $C$ represents the total number of known classes (as in the case of their work, where it is equal to 9), while $C+1$ refers to the total number of unknown classes (which, in their work, is 5). The Pavia dataset, the Indian Pines dataset, and the Salinas Valley dataset are the datasets that are evaluated in their study. Apart from mapping error, the Openess metric \cite{geng2020recent} is also considered as a benchmark for their evaluation, which evaluates the degree of openness for a given dataset in open-world classification, in addition to OA and micro F1 score. 

Equation \ref{eq13} elucidates the association between Openess and the number of training and testing data, $N_{train}$ and $N_{test}$, respectively.

\begin{equation} \label{eq13}
\begin{split}
Openess = 1 - \sqrt{\frac{2 \times N_{train}}{N_{train}+N_{test}}}.
\end{split}
\end{equation} 

\begin{figure}[hbt!]
    \centering
    \includegraphics[width=9.0cm, height = 6.5cm]{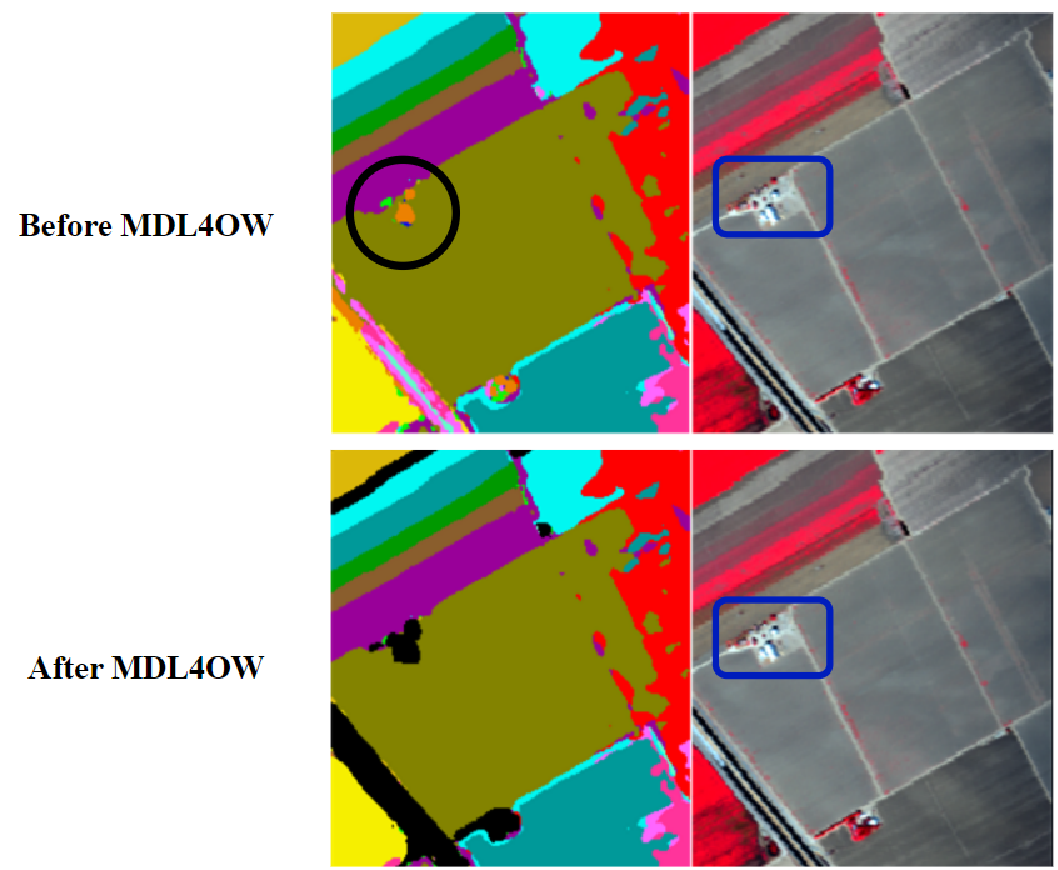}
    \caption{Illustration of how the MDL4OW can help to identify unknown semantic labels (The road highlighted in black and the house enclosed in the dark blue border) that the training network has not being exposed to beforehand. Any label that the network has not learned during training would be highlighted in black in the MDL4OW scheme (\textbf{Top}), as compared to without it, where the house is labelled in orange (enclosed by black circle). This may cause confusion as it may be misinterpreted as other classes (\textbf{Bottom}). Image is from the Salinas Valley dataset.}
\end{figure}

Figure 5 presents an illustration of how the MDL4OW methodology effectively identifies unknown classes, as demonstrated through an example image. The top portion of the figure highlights the road (denoted by black) and the house (enclosed by a dark-blue border), both of which cannot be assigned to any known class, as they were not presented \emph{a priori}. However, a standard deep learning model would still require assigning them a label, as it was not trained to recognize these specific labels. In contrast, the MDL4OW approach (depicted in the bottom portion of the figure) is adept at identifying and marking unknown classes (denoted by black), effectively applying the proposed scheme.

Employing adaptive subspace learning and feature-wise transformation (SSFT) techniques, \cite{bai2022few} aimed to enhance feature diversities and minimize overfitting. In particular, they incorporated a 3-D local channel attention residual network to extract features, and evaluated their algorithm against the SOTA using the Salinas, Pavia, and Indian Pines datasets. To compare with other SOTA, they performed a 5-shot, 10-shot, 15-shot, 20-shot, and 25-shot evaluation approach. In the study conducted by \cite{ding2021boosting}, a pseudo-labelling approach was adopted to augment the feature extraction procedure of their network using limited samples and also reduce overfitting. The soft pseudo label was computed by taking into account the euclidean distance between the unlabelled samples and the other agents with each labelled sample acting as a reference. Two sub-networks, namely the 3D-CNN and the SSRN (based on the ResNet), were proposed to function as the feature extractor. The dataset used for evaluation comprised of Pavia, Indian Pines and Salinas Valley. For comparison with other SOTA approaches, a 1-shot, 3-shot, and 5-shot evaluation approach was employed. Results showed that the proposed model outperformed all existing SOTA approaches in all three evaluation settings. 

Using a 3D residual convolutional block attention network (R3CBAM), the authors of \cite{pal2022few} demonstrated how to effectively learn spectral-spatial features in a more salient manner with small training samples. The CBAM is incorporated as an attention network. Meta-learning is employed, where a set of Euclidean distances from the test query set from known class prototypes are leveraged, and unknown class queries are labelled as outliers and are recognized without setting a threshold value beforehand. The evaluation of their approach was performed on the Indian Pines, Pavia, Salians, and Houston datasets. During the training of their network, a query set was generated from six base classes randomly chosen, and the support set was formed from samples using three randomly chosen query classes. A 1-shot and 5-shot OSR performance evaluation were carried out and compared with SOTA methods. In both 1-shot and 5-shot OSR evaluations, the results indicated that the proposed method outperformed the SOTA methods.

Expanding upon previous works, \cite{wang2021heterogeneous} proposed the Heterogeneous Few-Shot Learning (HFSL) approach for remote sensing classification with few samples per class. The method initially learns from data randomly sampled from the mini-ImageNet dataset to obtain transferable knowledge, followed by separating the data into support and query sets. A spectral-spatial fusion few-shot learning model is proposed that extracts spectral information through 1D mathematical operations and spatial information through a CNN with VGG16 pre-trained weights in the first layer. Their evaluation approach includes the Pavia University and Houston datasets, with a 5-shot performance evaluation against state-of-the-art methods. Building on this, \cite{hu2022heterogeneous} adds Knowledge Distillation (KD) to the approach, making it simpler to identify important parts of small samples, even with a shallower network. Further knowledge transfer and fine-tuning of the classifier model are performed, with evaluation on the Pavia University and Indian Pines datasets.

Using the Dirichlet-Net for feature extraction, \cite{qu2019few} suggested a few-shot multi-task transfer learning strategy that aims to maintain classification accuracy across several domains. The key concept is to extract fundamental representations that are common to the same type of object features across domains, with the aim of circumventing the requirement for more ground-truth labels from the target domain. The Pavia University dataset was employed to assess their approach, with a 5-sample per class evaluation strategy (i.e., 5-shot evaluation). Results showed that the proposed method was able to accurately classify unseen target domain samples, demonstrating the efficacy of the approach. 

The authors of \cite{tong2020few} proposed a new Attention-weight Graph Convolutional Network (AwGCN) for a few-shot method of quantifying and correlating internal features in hyperspectral data. This is followed by a semi-supervised label propagation of the node labels (features) from the support to query set via the GCN using the trained weights of the attention graphs. Unlike other approaches, they did not rely on pre-trained CNN-based weights as feature extractors but instead utilized a graph-based approach. The proposed method was evaluated on the Indian Pines dataset using 1-shot, 3-shot, and 5-shot approaches and on the Pavia University dataset using a 5-shot approach. In a similar vein, \cite{yang2020hyperspectral} proposed a GraphSAGE-based approach that utilizes spectral and spatial feature information to greatly reduce algorithmic space and time complexity. Their approach was evaluated on the Pavia University, Indian Pines, and Kennedy Space Centre dataset using a 30-samples per class evaluation approach for training and a 15-samples per class evaluation approach for validation. Their results demonstrated improved accuracy compared to other state-of-the-art methods, with a 6.7\% increase in accuracy on the Pavia University dataset and a 5.2\% increase on the Indian Pines dataset. Furthermore, the Kennedy Space Centre dataset improved accuracy by 7.1\%, making their approach a strong contender for further research. This improvement in accuracy indicates that their approach is more effective than other methods and could be a potential solution for future applications. 

The proposed method by \cite{huang2021self}, Self-Attention and Mutual-Attention Few-Shot Learning (SMA-FSL), utilizes a 3D convolutional feature embedding network for spectral-spatial extraction, coupled with a self-attention module to extract prototypes from each class in the support set and a mutual-attention module that updates and aligns these category prototypes with the query set. Attention-based learning is emphasized, where crucial features are enhanced while noisy features are reduced. To assess the efficacy of their approach, it is evaluated on the Houston, Botswana \cite{gerke2014use}, Chikusei \cite{yokoya2016airborne} and Kennedy Space Center datasets using 1-shot, 5-shot, and 15-shot evaluation approaches. These datasets are chosen for their diverse range of terrain and vegetation, allowing for a comprehensive evaluation of the model's performance. The results of the evaluation show that the approach is effective across all datasets, demonstrating its versatility in different environments.

The proposed work by \cite{zhao2022few} presents an incremental learning-based method that constantly updates the classifier by utilizing few-shot samples, allowing recognition of new classes while retaining knowledge of previous classes. The feature extractor module is implemented using a 20-layer ResNet, and the few-shot class incremental learning (FSCIL) is carried out via a constantly updated classifier (CUC), which is further enhanced by incorporating an attention mechanism for measuring the prototype similarity between each training and test sample class. The Pavia University dataset was employed for evaluating the performance of this approach using a 5-shot evaluation strategy. The results obtained showed that the proposed FSCIL with CUC and attention mechanism achieved superior performance compared to the baseline method. Furthermore, it was also observed that the performance improved with an increase in the number of shots.

Most previous works have employed CNN-based architectures for few-shot learning in hyperspectral image classification. However, CNNs can struggle with modeling long-range dependencies in spectral-spatial data when training samples are scarce. This has motivated recent interest in transformer architectures as an alternative.

In a notable contribution, Shi et al. \cite{bai2022generative} addressed the challenge of performance degradation observed in hyperspectral image classification methods when only a limited number of labeled samples are available for training. They proposed a unified framework with a Transformer Encoder and Convolutional Blocks to enhance feature extraction without needing extra data. The Transformer Encoder provides global receptive fields to capture long-range dependencies, while the Convolutional Blocks model local relationships. Their method achieved state-of-the-art results on few-shot hyperspectral tasks using public datasets, demonstrating the potential of transformers to advance few-shot learning in this domain.

Huang et al. \cite{huang2023hfc} also recognized limitations of CNN-based models for few-shot hyperspectral image classification. They highlighted the inherent difficulty of CNNs in effectively capturing long-range spatial-spectral dependencies, especially in scenarios with limited training data. They proposed an improved spatial-spectral transformer (HFC-SST) to overcome this, inspired by transformers' strong modeling capabilities for long-range relationships. HFC-SST generates local spatial-spectral sequences as input based on correlation analysis between spectral bands and adjacent pixels. A transformer-based network then extracts discriminative spatial-spectral features from this sequence using only a few labeled samples. Experiments on multiple datasets demonstrated that HFC-SST outperforms CNNs and prior few-shot learning methods by effectively modeling local long-range dependencies in limited training data. This further highlights the potential of transformers to advance few-shot hyperspectral classification through robust spatial-spectral feature learning.

The work by Wang et al. \cite{peng2023convolutional} also explores cross-domain few-shot learning for hyperspectral image classification, where labeled samples in the target domain are scarce. They propose a convolutional transformer-based few-shot learning (CTFSL) approach within a meta-learning framework. Most prior cross-domain few-shot methods rely on CNNs to extract statistical features, which only capture local spatial information. To address this, CTFSL incorporates a convolutional transformer network to extract both local and global features. A domain aligner maps the source and target domains to the same space, while a discriminator reduces domain shift and distinguishes feature origins. By combining few-shot learning across domains, transformer-based feature extraction, and domain alignment, their method outperforms state-of-the-art techniques on public hyperspectral datasets. This demonstrates the potential of transformers and cross-domain learning strategies to advance few-shot hyperspectral classification with limited labeled data.

Recently, Ran et al. \cite{ran2023deep} proposed a novel deep transformer and few-shot learning (DTFSL) framework for hyperspectral image classification that aims to overcome the limitations of CNNs. The DTFSL incorporates spatial attention and spectral query modules to capture long-range dependencies between non-local spatial samples. This helps reduce uncertainty and better represent underlying spectral-spatial features with limited training data. The network is trained using episode and task-based strategies to learn an adaptive metric space for few-shot classification. Domain adaptation is also integrated to align distributions and reduce variation across domains. Experiments on three public HSI datasets demonstrated that the transformer-based DTFSL approach outperforms state-of-the-art methods by effectively modeling relationships between non-local spatial samples in a few-shot context. This indicates transformers could be a promising alternative to CNNs for few-shot hyperspectral classification. 

In another work, \cite{liu2022feedback} introduces a Vision Transformer (ViT)-based architecture for FSL that employs feedback learning. The Few-Shot Transformer Network (FFTN) developed by \cite{liu2022feedback} combines spatial and spectral attention of the extracted features learned by the transformer component. By incorporating XAI, the model's decision-making process can be made more transparent and interpretable, thereby enhancing its trustworthiness and reducing the risk of biases. Additionally, the network incorporates meta-learning with reinforced feedback learning on the source set as the first step to improving the network's ability to identify misclassified samples through reinforcement. The second step is target-learning with transductive feedback training on the target sample to learn the distribution of unlabeled samples. This two-step process helps the network adapt to the target domain, thus improving its accuracy and reducing the risk of overfitting. 

The table \ref{tab3} provides an overview of some of the existing methods for few-shot approaches in hyperspectral image classification. It lists the dataset and metrics used, as well as the type of feature extractor approach for each method and the year of publication. The addition of XAI techniques could enhance the transparency and interpretability of these methods.

\begin{table}
\caption{Overview of few-shot hyperspectral image classification methods with feature extraction approach, publication year, and whether the method incorporates XAI techniques.\label{tab3}}
\resizebox{\textwidth}{!}{%

		\begin{tabular}{p{3cm}p{6cm}p{3cm}p{3cm}p{2cm}p{1cm}}  % Change this.
		\hline
			\textbf{Methods} & \textbf{Datasets} & \textbf{Metrics} & \textbf{Feature Extractor Approach} & \textbf{Publication Year} & \textbf{XAI} \\\hline
		
  MDL4OW \cite{liu2020few} & Pavia, Indian Pines, Salinas & OA, AA, $\kappa$ & Convolutional-Based & 2021 & No\tabularnewline
SSFT \cite{bai2022few} & Pavia, Indian Pines, Salinas & OA, AA, $\kappa$ & Convolutional-Based & 2022 & No\tabularnewline
3D-CNN, SSRN \cite{ding2021boosting} & Pavia, Indian Pines, Salinas & OA, AA, $\kappa$ & Convolutional-Based & 2021 & No\tabularnewline
R3CBAM \cite{pal2022few} & Pavia, Indian Pines, Salinas, Houston & OA, AA, $\kappa$ & Convolutional-Based & 2022 & No\tabularnewline
HFSL \cite{wang2021heterogeneous} & Pavia, Houston & OA, AA, $\kappa$ & Convolutional-Based & 2021 & No\tabularnewline
HFSL + KD \cite{hu2022heterogeneous} & Pavia, Indian Pines & OA, AA, $\kappa$ & Convolutional-Based & 2022 & No\tabularnewline
AwGCN \cite{tong2020few} & Pavia, Indian Pines & OA, AA, $\kappa$ & Graph-Based & 2020 & No\tabularnewline
GraphSAGE \cite{yang2020hyperspectral} & Pavia, Indian Pines, Kennedy Space Centre & OA, AA, $\kappa$ & Graph-Based & 2021 & No\tabularnewline
Multi-task FSL \cite{qu2019few} & Pavia & OA, AA, $\kappa$ & Convolutional-Based & 2019 & No\tabularnewline
SMA-FSL \cite{huang2021self} & Houston, Botswana, Chikusei, Kennedy Space Centre & OA, AA, $\kappa$ & Convolutional-Based & 2021 & No\tabularnewline
FSCIL \cite{zhao2022few} & Pavia & OA, AA, $\kappa$ & Convolutional-based (Incremental) & 2022 & No\tabularnewline
FFTN \cite{liu2022feedback} & Pavia, Chiksuei, Salinas Valley & OA, AA, $\kappa$ & Vision Transformers & 2022 & Yes\tabularnewline
	\hline
		\end{tabular}}

\end{table}

A 5-shot evaluation method is mostly used to measure how well the proposed methods work on the Chikusei, Salinas Valley, and Pavia University datasets. Table \ref{tab3} gives an overview of the methods that have been discussed, including the datasets and evaluation metrics, the learning methods that were used, and the year the paper was published. From the works that were talked about, it is clear that the Pavia, Indian Pines, and Salinas Valley datasets are the ones most often used to compare algorithms. 

In the table \ref{tab3}, most of the few-shot approaches for hyperspectral image classification have not incorporated XAI for better interpretability. However, the addition of XAI techniques could enhance transparency and provide insights into the decision-making process of the models. One way to incorporate XAI is by using visualization techniques to highlight the features or regions of the image that contribute to the model's prediction. Another approach is to use saliency maps to identify the most important regions of the input image that influence the model's decision. Additionally, model-agnostic methods such as LIME or SHAP can provide insights into the decision-making process of the models. Overall, the incorporation of XAI techniques in few-shot approaches for hyperspectral image classification can improve the transparency and interpretability of the models and facilitate their adoption in real-world applications.

\subsection{Few-shot learning in VHR image classification} 

All studies in this section employ the OA metric unless otherwise stated. \cite{li2017zero} proposes a novel zero-shot scheme for scene classification (ZSSC) based on the visual similarity of images from the same class. Their work employs the UC Merced dataset for evaluation, where a number of classes were randomly chosen as observed classes while the rest were unseen classes. Additionally, the authors incorporate the RSSCN7 dataset \cite{zou2015deep} and a VHR satellite-based image database consisting of instances from both observed and unknown classes in an unlabeled format. To address the issue, the authors adopt the word2vec \cite{church2017word2vec} model to represent each class as a semantic vector and use a K-Nearest Neighbor (KNN) graph-based model to implement sparse learning for label refinement. The refinement also helps to denoise any noisy label during the zero-shot classification scheme. Their proposed model achieves significant performance gains compared to existing SOTA zero-shot learning models with linear computational complexity. Furthermore, the proposed model can handle a large number of classes with minimal memory requirements.Some of the newest attempts to use few-shot learning and UAV-based data was carried out by \cite{al2023intelligent,khoshboresh2023multimodal}. 

The work presented by \cite{hamzaoui2022hierarchical} proposed a hierarchical prototypical network (HPN) as a novel approach for few-shot learning, which is evaluated on the RESISC45 dataset. The HPN model is designed to perform analysis of high-level aggregated information in the image, followed by fine-level aggregated information computation and prediction, utilizing prototypes associated with each level of the hierarchy as described in (4). The evaluation protocol involves a 5-way 1-shot and 5-way 5-shot classification approach within the standard meta-learning framework. In the proposed approach, the ResNet-12 model serves as the backbone for the first stage of feature extraction. The extracted features are then passed through a linear layer to obtain the final feature representation. This feature representation is used for the classification task in the second stage. 

In a bid to augment the performance of few-shot task-specific contrastive learning (TSC), \cite{zeng2022task} introduced a self-attention and mutual-attention module (SMAM) that scrutinizes feature correlations with the aim of reducing any background interference. The adoption of a contrastive learning strategy facilitates the pairing of data using original images from diverse perspectives. Ultimately, the aforementioned approach enhances the potentiality to distinguish intra-class and inter-class image features. The NWPU-RESISC45, WHU-RS19, and UC Merced datasets were leveraged for their algorithmic evaluation, which comprised a 5-way 1-shot and 5-way 5-shot classification approach that was scrutinized and compared. Furthermore, \cite{yuan2020few} introduced a multiple-attention approach that concurrently focuses on the global and local feature scale as part of their Multi-Attention Deep Earth Mover Distance (MAEMD) proposed network. Local attention is geared towards capturing significant and subtle local features while suppressing others, thereby improving representational learning performance and mitigating small inter-class and large intra-class differences. Their approach was evaluated on the UC-Merced, AID, and OPTIMAL-31 \cite{wang2018scene} datasets with a 1-shot, 5-way 5-shot, and 10-shot evaluation approach. As evidence of the success of the local attention strategy, the results showed that the model achieved state-of-the-art performance across all datasets.

Another illustration of an attention-based model is presented by \cite{kim2021saffnet} with the introduction of the Self-Attention Feature Selection Network (SAFFNet). This model aims to integrate features across multiple scales using a self-attention module, in a similar manner to that of a spatial pyramid network. The Self-Attention Feature Selection (SAFS) module is employed to better match features from the query set with the fused features in the class-specified support set. Experimental analysis was conducted on the UC-Merced, RESISC45, and AID datasets, using a 1-shot and 5-shot classification evaluation approach. The results showed that SAFS was able to improve the performance of the baseline model for all datasets, with the largest improvement seen on the AID dataset.

Incorporating a feature encoder to learn the embedded features of input images as a pre-training step, \cite{huang2021taes} proposed the Task-Adaptive Embedding Network (TAE-Net). To choose the most informative embedded features during the learning task in an adaptive manner, a task-adaptive attention module is employed. By utilizing only limited support samples, the prediction is performed on the query set by the meta-trained network. For their algorithmic evaluation, they employed the NWPU-RESISC45, WHU-RS19, and UC Merced dataset. A 5-way 1-shot and 5-way 5-shot classification approach were implemented for comparison. The results were evaluated based on accuracy, precision, recall, and F1-score metrics. Furthermore, the models were compared in terms of their training time and memory usage.

The study conducted by \cite{wang2022few} aims to achieve few-shot learning through deep economic networks. The deep economic network incorporates a two-step simplification process to reduce training parameters and computational costs in deep neural networks. The reduction of redundancy in input image, channel, and spatial features in deep layers is achieved. In addition, teacher knowledge is utilized to improve classification with limited samples. The last block in the model includes depth- and point-wise convolutions that effectively learn cross-channel interactions and enhance computational efficiency. The algorithmic evaluation of the model is conducted on three datasets, namely the UC-Merced, RESISC45, and RSD46-WHU. The evaluation is carried out using a 1-shot and 5-shot approach on RESISC45 and the RSD46-WHU, and an additional 10-shot evaluation is implemented on the UC-Merced dataset. The model shows promising performance across all datasets, with the highest accuracy coming from the 10-shot evaluation.

The introduction of the Discriminative Learning of Adaptive Match Network (DLA-MatchNet) for few-shot classification by \cite{li2020dla} incorporated the attention mechanism in the channel and spatial domains to identify the discriminative feature regions in the images through the examination of their inter-channel and inter-spatial relations. In order to address the challenges posed by large intra-class variances and inter-class similarity, the discriminative features of both the support and query sets were concatenated, and the most relevant pairs of samples were adaptively selected by a matcher, which was manifested as a multi-layer perceptron. The UC-Merced, RESISC45, and WHU-RS19 datasets were employed for the state-of-the-art (SOTA) evaluation, utilizing a 5-way 1-shot and 5-shot approach for all datasets. The results confirmed the superior accuracy of the proposed method over the SOTA, proving its utility for remote sensing image retrieval. 

Graph-based methods have also been employed in the very high-resolution (VHR) domain for few-shot learning. In this regard, \cite{jiang2022multi} proposed a multi-scale graph-based feature fusion (MGFF) approach that involves a feature construction model that converts typical pixel-based features to graph-based features. Subsequently, a feature fusion model combines the graph features across several scales, which enhances the distinguishing ability of the model via integrating the essential semantic feature information and thereby improving few-shot classification capability. The authors conducted the algorithmic evaluation on the RESISC45 and WHU-RS19 datasets using a 5-way 1-shot and 5-way 5-shot classification approach. In addition, \cite{yuan2022graph} proposed Graph Embedding Smoothness Network (GES-Net), which implements embedded smoothing to regularize the embedded features. This not only effectively extracts higher-order feature relations but also introduces a task-level relational representation that captures graph relations among the nodes at the level of the whole task, thereby enhancing the node relations and feature discerning capabilities of the network. The work is evaluated on the RESISC45, WHU-RS19, and UC Merced datasets using 5-way 1-shot and 5-shot comparison approaches. Episodic training was adopted, where each episode refers to a task and is comprised of $N$ uniformly sampled categories without replacement and the query and support set. The support set contains $K$ samples from each of the $N$ categories, and the query set contains a single sample from each of the $N$ categories. The samples in the query and support sets are selected from a larger pool of available samples in a random manner, ensuring that each episode is unique. 

Few-shot strategies for VHR image classification can benefit greatly from the application of XAI methods like explainable graph neural networks and attention mechanisms. Transparency, accountability, bias detection, and fairness issues can all be improved with the help of xGNNs because they shed light on the model's decision-making process. To make the model's decisions more understandable and transparent, attention mechanisms can draw focus to key features and nodes in the graph. In principle, these methods could make graph-based few-shot classification models more reliable and easy to understand.

Table \ref{tab4} presents a similar format to Table \ref{tab3}, depicting a comprehensive summary of the aforementioned methods in the very high-resolution (VHR) classification domain. It can be observed that the RESISC45, UC-Merced, and WHU-RS19 dataset are among the most frequently employed for algorithmic comparison, as seen in the subset of the existing works highlighted.

\begin{table}
\caption{The overview of some existing methods for few-shot approaches in VHR image classification. In addition to the dataset and metrics used, we also include the type of feature extractor approach for method and the year at which the work was published.\label{tab4}}
\resizebox{\textwidth}{!}{%
		\begin{tabular}{p{5cm}p{5cm}p{1cm}p{3cm}p{2cm}}  % Change this.
			\hline
			\textbf{Methods} & \textbf{Datasets} & \textbf{Metrics} & \textbf{Feature Extractor Approach} & \textbf{Publication Year}\\\hline
		
  ZSSC \cite{li2017zero} & UC Merced, RSSCN7 & OA & Sparse Learning & 2017 \tabularnewline
  HPN \cite{hamzaoui2022hierarchical} & RESISC45 & OA & Convolutional-Based & 2022 \tabularnewline
  SMAM \cite{zeng2022task} & RESISC45, WHU-RS19, UC Merced & OA  & Convolutional-based & 2022 \tabularnewline
  MAEMD \cite{yuan2020few} & RESISC45, AID, OPTIMAL-31 & OA  & Convolutional-based & 2021 \tabularnewline
  SAFFNet \cite{kim2021saffnet} & RESISC45, AID, UC Merced & OA & Convolutional-Based & 2021 \tabularnewline
  TAE-Net \cite{huang2021taes} & RESISC45, WHU-RS19, UC Merced & OA & Convolutional-based & 2021 \tabularnewline
  Deep Economic Network \cite{wang2022few} & RESISC45, UC-Merced, SD46-WHU & OA & Convolutional-Based & 2022 \tabularnewline
  DLA-MatchNet \cite{li2020dla} & RESISC45, WHU-RS19, UC Merced & OA & Convolutional-Based & 2021 \tabularnewline
  MGFF \cite{jiang2022multi} & RESISC45, WHU-RS19 & OA & Graph-Based & 2022 \tabularnewline
  GES-Net \cite{yuan2022graph} & RESISC45, WHU-RS19, UC Merced & OA & Graph-Based & 2022 \tabularnewline\hline
		\end{tabular}}
\end{table}

Furthermore, the majority of the approaches presented in Table 4 incorporate attention mechanisms and graph-based methods for few-shot VHR classification. The MGFF approach presented by \cite{jiang2022multi} is an example of a graph-based method, while the DLA-MatchNet presented by \cite{li2020dla} is an example of an approach that utilizes attention mechanisms. Similarly, the GES-Net presented by \cite{yuan2022graph} also uses graph-based methods in its approach. These techniques aim to extract more relevant and informative features from the input images, which can enhance the performance of few-shot VHR classification systems.

\subsection{Few-shot learning in SAR image classification}

SAR (Synthetic Aperture Radar) is a remote sensing technology for capturing high-resolution images of the Earth's surface regardless of the weather conditions, making it a valuable tool in various applications such as agriculture, forestry, and land use management. However, the availability of SAR-based data is often limited in comparison to hyperspectral or VHR-based data, mainly due to the high cost of SAR sensors and the complexity of SAR data processing. As a result, traditional classification approaches for SAR data are often challenged by insufficient training data and the high intra-class variability, which leads to a pressing need for the development of few-shot learning methods that can effectively tackle these challenges. Therefore, a review of emerging few-shot learning methods in the SAR-based classification domain is highly desirable to advance the state-of-the-art and enable more accurate and efficient classification of SAR data.

The integration of XAI techniques, such as explainable graph neural networks (xGNNs) and attention mechanisms, can significantly enhance the proposed few-shot transfer learning technique for SAR image classification presented by \cite{tai2022few}. This novel approach uses a connection-free attention module to selectively transfer shared features between SAR and Electro-Optical (EO) image domains, reducing the dependence on additional SAR samples, which may not be feasible in certain scenarios. By using xGNNs, the authors can provide insights into the decision-making process, increasing transparency and accountability, which is particularly crucial for SAR image classification due to restricted data access and high acquisition costs. The attention mechanism can highlight relevant features and nodes in the graph, improving the model's interpretability and transparency, and ultimately, its performance and trustworthiness. In addition, the authors implemented a Bayesian convolutional neural network to update only relevant parameters and discard those with high uncertainties. The evaluation was performed on three EO datasets that included ships, planes, and cars, with SAR images obtained from \cite{hammell2019data}, \cite{schwegmann2017sar}, and MSTAR. The classification accuracy (OA) value was used as the performance metric, with the 10-way $k$-shot approach achieving an OA of approximately 70\%, outperforming other approaches. Overall, incorporating XAI techniques can potentially improve the performance and trustworthiness of the proposed few-shot transfer learning technique for SAR image classification.

In the pursuit of effective few-shot classification, \cite{gao2022few} proposed a Hand-crafted Feature Insertion Module (HcFIM), which combines learned features from CNN with hand-crafted features via a weighted-concatenated approach to aggregate more priori knowledge. Their Multi-scale Feature Fusion Module (MsFFM) is used to aggregate information from different layers and scales, which helps distinguish target samples from the same class more easily. The combination of MsFFM and HcFIM forms their proposed Multi-Feature Fusion Network (MFFN). To tackle the challenge of high similarity within inter-classes in SAR images, the authors proposed the Weighted Distance Classifier (WDC), which computes class-specific weights for query samples in a data-driven manner, distributed using the Euclidean distance as a guide. They also incorporated weight generation loss to guide the process of weight generation. The benchmark MSTAR dataset and their proposed Vehicles and Aircraft (VA) dataset were used for evaluation, where a 4-way 5-shot evaluation approach was used for MSTAR and a 4-way 1-shot evaluation approach was used for VA. The Average Accuracy (AA) was used as the evaluation metric throughout. The evaluation results demonstrated that the VA dataset had a higher AA than MSTAR, indicating that it was better suited for fine-grained classification tasks. 

% The implementation of a meta-learning approach in few-shot classification is proposed by \cite{fu2021few}, utilizing a meta-learner and a base-learner. The meta-learner stores the learning rates and generalized parameters for the feature extractor and classifier to learn optimal initialization parameter and update strategies through analyzing the distribution of few-shot tasks. Subsequently, the meta-learner guides the base-learner, which is the SAR-based target classifier model, to converge more efficiently. To tackle harder tasks, a hard-task mining technique is incorporated. During the meta-training phase, the 4CONV network is used to acquire transferrable knowledge. The evaluation is conducted on the MSTAR dataset and the proposed NIST-SAR, utilizing a 5-way 1-shot and 5-way 5-shot approach, with the mean accuracy (AA) being used as the evaluation metric. The approach is referred to as MSAR in the paper. The results demonstrate that the proposed MSAR approach outperforms the baseline methods on both tasks, achieving an AA of 86.2\% for the 5-way 1-shot task and 97.5\% for the 5-way 5-shot task. 

 In the study by \cite{fu2021few}, a novel approach to few-shot classification is introduced through the integration of meta-learning. This method is characterized by the synergistic use of two primary components: a meta-learner and a base-learner. The meta-learner's primary function is to determine and store the learning rates, along with generalized parameters pertinent to both the feature extractor and classifier. Its objective is to discern an optimal initialization parameter, thereby refining update strategies by meticulously examining the distribution of few-shot tasks. This optimal initialization is instrumental in setting the algorithm on a path that potentially accelerates convergence and improves performance. Following this, the meta-learner plays a pivotal role in directing the base-learner. Here, the base-learner is conceptualized as a classifier model specifically tailored for SAR-based target detection. Its design ensures enhanced convergence efficiency under the guidance of the meta-learner.

Recognizing the challenges posed by more complex tasks, the study further augments its methodology with a hard-task mining technique. This is particularly valuable in emphasizing and addressing tasks that are inherently more challenging. For the acquisition of transferrable knowledge—a crucial aspect of few-shot learning—the 4CONV network is employed during the meta-training phase. The efficacy of this approach, termed as MSAR in the publication, was rigorously tested on two datasets: the MSTAR dataset and the newly proposed NIST-SAR dataset. Evaluations were carried out using both the 5-way 1-shot and 5-way 5-shot paradigms, with the mean accuracy (AA) serving as the benchmark metric. The empirical results were telling; the MSAR method surpassed baseline methodologies in performance for both tasks. Specifically, it achieved an impressive AA of 86.2\% for the 5-way 1-shot task and an even more commendable 97.5\% for the 5-way 5-shot task.

The paper by \cite{rostami2019sar} proposed a novel few-shot cross-domain transfer learning approach to transfer knowledge from the electro-optical (EO) domain to the synthetic aperture radar (SAR) domain. This is accomplished by utilizing an encoder in each domain to extract and embed individual features into a shared embedded space. The encoded parameters are updated continuously by minimizing the discrepancies in the marginal probability distributions between the two embedded domains. Since the distributions are generally unknown in few-shot learning, the authors approximate the optimal transport discrepancy measurement metric using the Sliced Wasserstein Distance (SWD) for more efficient computation. The approach is evaluated on a dataset of SAR images acquired by \cite{schwegmann2016very} for detecting the presence or absence of ships. The classification accuracy (OA) is used as the evaluation metric for this approach. The results show that this approach can achieve an OA of over 90\%, indicating that it is a reliable and accurate method for ship detection.

In their study on few-shot ship recognition using the MSTAR dataset, \cite{wang2022fewDKL} proposed a Deep Kernel Learning (DKL) approach that harnesses the non-parametric adaptability of Gaussian Processes (GP). The kernel function used in their approach is mathematically defined in (14) as a Gaussian kernel

\begin{equation}
k_{l}(x,\bar{x}) = exp\left(\frac{-\norm{x-\bar{x}}^{2}}{2l^{2}}\right)
\end{equation}

which determines the similarity and relationships between pairs of embedded data samples $x$ and $\bar{x}$, with $l$ serving as a hyperparameter characterizing the length scale. The DKL approach integrates such kernel functions with deep neural networks to enable effective few-shot classification. For the $K$-shot $C$-way few-shot classification, the authors trained GPs on $C$ categories, where the \emph{ith} GP is trained on positive samples from class $C_{i}$ and negative samples from the remaining classes $C-1$. The GP with the highest confidence in the correct target classes is then computed using the log-likelihood formula. They evaluated their approach using 1-shot and 5-shot classification, with the classification accuracy (OA) as the metric.

Graph-based learning methods have gained popularity in SAR image classification similar to hyperspectral and VHR image classification. To enhance feature similarity learning among query images and support samples more effectively using graphs, \cite{yang2020learning} proposed a relation network based on the embedding network for feature extraction and attention-based Graph Neural Networks (GNN) in the form of a metric network \cite{sung2018learning}. The channel attention module in CBAM is incorporated into the GNN. MSTAR is utilized for evaluation, and a 5-way 1-shot comparison is used with classification accuracy (OA) as the metric. In addition, Yang proposed a Mixed-loss Graph Attention Network (MGA-Net) which utilizes a multi-layer GAT combined with a mixed-loss (embedding loss and classification loss) training to increase inter-class separability and speed up convergence. The MSTAR dataset and the OpenSARShip dataset were used for comparison, and a 3-way 1-shot and 3-way 5-shot classification evaluation were utilized for comparison of the results represented by the classification accuracy (OA) and the confusion matrix. The results showed that the MGA-Net achieved a better performance than the baseline models in both datasets, indicating that the multi-layer GAT and mixed-loss training had a positive effect on the classification accuracy.

Recently, Zhao et al. \cite{zhao2022few2} proposed an instance-aware transformer (IAT) model for few-shot synthetic aperture radar automatic target recognition (SAR-ATR). They recognize that modeling relationships between query and support images is critical for few-shot SAR-ATR. The IAT leverages transformers and attention to aggregate relevant support features for each query image. It constructs attention maps based on similarities between query and support features to exploit information from all instances. Shared cross-transformer modules align query and support features. Instance cosine distance during training pulls same-class instances closer to improve compactness. Experiments on few-shot SAR-ATR datasets show IAT outperforms state-of-the-art methods. Visualizations also demonstrate improved intra-class compactness and inter-class separation. This highlights the potential of transformers and attention for few-shot SAR classification by effectively relating queries to supports and learning discriminative alignments.

CNNs have been dominant for SAR-ATR, but struggle with limited training data. To address this, Wang et al. \cite{wang2022global} proposed a convolutional transformer (ConvT) architecture tailored for few-shot SAR ATR. They recognize that CNNs are hindered by narrow receptive fields and inability to capture global dependencies in few-shot scenarios. ConvT constructs hierarchical features and models global relationships of local features at each layer for more robust representation. A hybrid loss function based on recognition labels and contrastive image pairs provides sufficient supervision from limited data. Auto augmentation further enhances diversity while reducing overfitting. Without needing additional datasets, ConvT achieves state-of-the-art few-shot SAR ATR performance on MSTAR by effectively combining transformers with CNNs. This demonstrates transformers can overcome CNN limitations for few-shot SAR classification by integrating local and global dependencies within and across layers.

\begin{table}
\caption{The overview of some existing methods for few-shot approaches in SAR image classification. In addition to the dataset and metrics used, we also include the type of feature extractor approach for method and the year in which the work was published.\label{tab5}}
\resizebox{\textwidth}{!}{%
		\begin{tabular}{p{5cm}p{4cm}p{1cm}p{3cm}p{3cm}}  % Change this.
\hline
			\textbf{Methods} & \textbf{Datasets} & \textbf{Metrics} & \textbf{Feature Extractor Approach} & \textbf{Publication Year}\\\hline
		
 Bayesian-CNN \cite{tai2022few} & \cite{hammell2019data}, \cite{schwegmann2017sar}, MSTAR & OA & Convolutional-Based & 2022 \tabularnewline
 MFFN +WDC \cite{gao2022few} & MSTAR, VA & AA & Convolutional-Based & 2022 \tabularnewline
 MSAR \cite{fu2021few} & MSTAR, NIST-SAR & AA & Meta-Learning & 2021 \tabularnewline
 Cross-Domain Transfer \cite{rostami2019sar} & \cite{schwegmann2016very} & OA & Transfer Learning & 2019 \tabularnewline
 DKL \cite{wang2022fewDKL} & MSTAR & OA & Kernel Learning & 2022 \tabularnewline
 GNN-based Relation Network \cite{yang2020learning} & MSTAR & OA & Graph-Based & 2020 \tabularnewline
 MGA-Net \cite{yang2021mixed} & MSTAR, OpenSARShip  & OA & Graph-Based & 2021 \tabularnewline
 IAT \cite{zhao2022few2} & MSTAR  & OA & Cross-transformer module & 2022 \tabularnewline
 ConvT \cite{wang2022global} & MSTAR  & AA & Convolutional transformer & 2022 \tabularnewline
 
 \hline
			
		\end{tabular}}
\end{table}

Table \ref{tab5} provides an overview of the various few-shot learning methods that have been proposed for SAR classification. The table summarizes the key aspects of each approach, including the name of the method, the year of publication, the dataset used for evaluation, and the evaluation metric used. It is noteworthy that among the subset of existing works described in this review, the MSTAR dataset is the most commonly used for algorithmic comparisons. The MSTAR dataset has been widely used in SAR classification due to its relatively large size and the diversity of the target types that it contains. Overall, the methods discussed in Table 5 highlight the potential of few-shot learning approaches in the SAR domain, and demonstrate the effectiveness of various techniques such as graph-based learning, deep kernel learning, and meta-learning. These approaches have the potential to enable more efficient and accurate classification of SAR data, which can have important applications in fields such as remote sensing, surveillance, and defense. XAI techniques can be really useful for identifying objects in radar images. Because researchers usually have limited access to radar data and it's expensive to get new radar images, techniques like explainable graph neural networks and attention mechanisms are helpful.

\section{Few-shot based object detection and segmentation in remote sensing}
In the remote sensing domain, much of the focus has been on image classification tasks like land cover mapping. However, it is also essential to advance higher-level vision tasks like object detection and semantic segmentation, which extract richer information from imagery. For example, object detection can precisely localize and identify vehicles, buildings, and other entities within a scene. Meanwhile, segmentation can delineate land, vegetation, infrastructure, and water boundaries at the pixel level. Significant progress has been made in developing and evaluating object detection and segmentation techniques for remote sensing data. Various benchmarks and competitions have been organized using large-scale satellite and aerial datasets \cite{li2020object, han2018advanced}. State-of-the-art deep learning models like R-CNNs, SSDs, and Mask R-CNNs \cite{su2019object} have shown strong performance. However, many of these rely on extensive annotated training data which can be costly and time-consuming to collect across diverse geographical areas. Therefore, advancing object detection and segmentation in remote sensing using limited supervision remains an open challenge. Few-shot learning offers a promising approach to enable effective generalization from scarce training examples. While some initial work has explored object detection for aerial images \cite{zhu2015orientation,yao2019unmanned,mundhenk2016large, zhang2019hierarchical, lu2019gated}, a comprehensive survey incorporating the latest advancements is still lacking. Furthermore, few-shot semantic segmentation has received relatively little attention for remote sensing thus far.
 
\subsection{Few-shot object detection in remote sensing}
The main challenge in few-shot object detection is to design a model that can generalize well from a small number of examples \cite{li2021few,wolf2021double, cheng2021prototype, gao2021fasts}. This is typically achieved by leveraging prior knowledge learned from a large number of examples from different classes (known as base classes). The model is then fine-tuned on a few examples from the new classes (known as novel classes) \cite{jeune2023rethinking}.

There are various methods used in few-shot object detection, including metric learning methods, meta-learning methods, and data augmentation methods \cite{xiao2021few, li2022mm, wang2022multi}. Metric learning methods aim to learn a distance function that can measure the similarity between objects. Meta-learning methods aim to learn a model that can quickly adapt to new tasks with a few training examples with the help other domain informations \cite{zhang2023text}. Data augmentation methods aim to generate more training examples by applying transformations to the existing examples \cite{liu2023transformation}. Furthermore, a more comprehensive analysis of aerial image-based FSOD is available in the summarized Table \ref{table:FSOD_aerial}. 

Explainability in few-shot object detection refers to the ability to understand and interpret the decisions made by the model. This is important for verifying the correctness of the model’s predictions and for gaining insights into the model’s behavior. Explainability can be achieved by visualizing the attention maps of the model, which show which parts of the image the model is focusing on when making a prediction. Other methods include saliency maps \cite{petsiuk2021black}, which highlight the most important pixels for a prediction, and decision trees, which provide a simple and interpretable representation of the model’s decision process \cite{hu2023xaitk}. Therefore, few-shot object detection methods have shown promising results in detecting novel objects in aerial images with limited annotated samples. The feature attention highlight module and the two-phase training scheme contribute to the model’s effectiveness and adaptability in few-shot scenarios. However, there are still challenges to be addressed, such as the performance discrepancy between aerial and natural images, and the confusion between some classes. Future research should focus on developing more versatile few-shot object detection techniques that can handle small, medium, and large objects effectively, and provide more interpretable and explainable results.

\begin{table}
\caption{Overview of studies addressing challenges in few-shot object detection for aerial remote sensing images }
\scalebox{0.7}{
\begin{tabular}{|l|l|l|l|l|}
\hline
\textbf{References} & \textbf{Challenges} & \textbf{Datasets used} & \textbf{Year} \\
\hline

%%%%% few shot object detection 2021
DH-FSDe \cite{wolf2021double} & Unseen class and limited samples  &  iSAID and NWPU VHR-10 & 2021 \\
FSODM \cite{li2021few} &  Generalization unseen class   &  DIOR,  NWPU VHR& 2021 \\
P-CNN \cite{cheng2021prototype} &  Robust and scarcity of novel class   &  DIOR & 2021 \\
Gao et all. \cite{gao2021fasts} &  Varying scale of novel class   &  NWPU VHR and DIOR & 2021 \\
SAAN \cite{xiao2021few} &  Varying scale of novel class   & RSOD and NWPU VHR  & 2021 \\
simple-CNNs \cite{xiao2021few} &  Resource-limited   & DIOR and NWPU VHR  & 2021 \\
PennSyn2Real \cite{nguyen2021pennsyn2real} &  Novel datasets introduced   &  PennSyn2Real  & 2021 \\
SAFFNet \cite{nguyen2021pennsyn2real} &  Varying scale of novel class   &  NWPU-RESISC45, AID and UCM  & 2021 \\
PRNet \cite{liu2021polar} & anchor-free novel class   &   DOTA and HRSC2016  & 2021 \\
FENet \cite{cheng2021feature} & Varying scale of novel class   &   DOTA and DIOR  & 2021 \\
%%%%%  2022
Jeune et all. \cite{le2022improving} &  benchmark analysis  &   DOTA and DIOR  & 2022 \\ 
MM-RCNN \cite{li2022mm} & Inter-class correlation between few samples  &   DIOR and NWPU VHR  & 2022 \\
DC-DML \cite{li2022aifs} & Generalization unseen class   &   AIFS-DATASET  & 2022 \\
MSSA \cite{wang2022multi} & Varying scale of novel class   &    NWPU VHR    & 2022 \\
MSSA \cite{su2022multi} & Generalization unseen class   &    RSOD    & 2022 \\
FSRDD \cite{su2022fsrdd} & Varying scale of novel class   &    RSOD    & 2022 \\

%%%%%  2023

TEMO \cite{lu2023few} & text-image fusion and limited varying samples  &  DIOR, NWPU and FAIR1M & 2023 \\
Wang et al. \cite{wang2023few} & Varying scale of novel class  &  NWPU VHR and DIOR & 2023 \\
Li et al. \cite{jeune2023rethinking} &  Robust of novel class  &  DOTA and DIOR & 2023 \\ 
Zhang et al. \cite{zhang2023text} & Multi-modal fusion for novel class   &  NWPU VHR and DIOR & 2023 \\
TINet \cite{liu2023transformation} & Varying scale of novel class   &  NWPU VHR, DIOR, and HRRSD & 2023 \\
BSFCDet \cite{liu2023transformation} & Generalization unseen class   &  SPPF-G, and SPP & 2023 \\ \hline
\end{tabular}}
\label{table:FSOD_aerial}
\end{table}

\subsection{FSOD benchmark datasets for aerial remote sensing images}

\begin{itemize}
    \item \textbf{NWPU VHR}  contains 10 categories, with three chosen as novel classes. Researchers commonly use a partition that involves base training on images without novel objects and fine-tuning on a set with  $\emph{k}$ annotated boxes (where $\emph{k}$ is 1, 2, 3, 5, or 10) for each novel class. The test set has about 300 images, each containing at least one novel object.
    \item \textbf{DIOR.} dataset features 20 classes and over 23,000 images. Five categories are designated as novel, and various few-shot learning approaches are applied. Fine-tuning is performed with $\emph{k}$ annotated boxes (where $\emph{k}$ can be 3, 5, 10, 20, or 30) for each novel class, and performance is evaluated on a comprehensive validation set.
    \item \textbf{RSOD.} dataset consists of four classes. One class is randomly selected as the novel class, with the remaining three as base classes. During base training, 60\% of samples for each base class are used for training, and the rest for testing. Fine-tuning is performed on $\emph{k}$ annotated boxes in the novel classes, where $\emph{k}$ can be 1, 2, 3, 5, or 10.
    \item \textbf{iSAID} dataset features 15 classes and employs three distinct base/novel splits designed according to data characteristics. Each split focuses on a different aspect—such as object size or variance in appearance. The third split specifically selects the six least frequent classes as novel. Base training uses all objects from base classes, and fine-tuning utilizes 10, 50, or 100 annotated boxes per class.
    \item \textbf{DOTA} dataset features an increase from 15 to 18 categories and nearly tenfold expansion to 1.79 million instances. It has two base/novel class splits, with three classes designated as novel. During episode construction, the number of shots for novel classes varies as 1, 3, 5, and 10.
    \item \textbf{DAN} dataset is an amalgamation of DOTA and NWPU VHR datasets, comprising 15 categories. It designates three classes as novel, with the remaining as base classes. 
\end{itemize}

\subsection{Few-shot image segmentation in remote sensing}
Few-shot image segmentation (FSIS) is a challenging task in computer vision, particularly in the context of aerial images. This task aims to segment objects in images with only a few labeled examples, which is especially important due to the high cost of collecting labeled data in the domain of aerial images. Recent advancements in few-shot image segmentation have been driven by deep learning techniques, which have shown promising results in various computer vision tasks. Metric-based meta-learning models, such as Siamese networks and prototype networks, have been widely used in few-shot segmentation \cite{yao2021scale, chen2022semi}. These models learn to compare the similarity between support and query images and use this information to segment novel classes\cite{cao2023few}.

Another common approach in few-shot image segmentation is to use deep learning networks, specifically convolutional neural networks (CNNs) \cite{zhang2020few}. These networks have shown great success in image segmentation tasks and have been adapted for few-shot learning scenarios. Researchers have explored different architectures and training strategies to improve the performance of CNNs in few-shot image segmentation\cite{zhang2020few}. Meta-learning, which involves training a model to learn how to learn, has also been applied to few-shot image segmentation with promising results \cite{zhang2020few}. Meta-learning algorithms aim to extract meta-knowledge from a set of tasks and use this knowledge to quickly adapt to new tasks with only a few labeled examples. In terms of applications, few-shot image segmentation in aerial images has various potential applications. One application is in urban planning, where few-shot image segmentation can be used to identify and segment different types of buildings, roads, and other urban infrastructure \cite{puthumanaillam2023texture,lang2023global,lang2023progressive}. Another application is in land-use and land-cover determination, where few-shot image segmentation can be used to classify different types of land cover, such as forests, agricultural land, and water bodies. Few-shot image segmentation can also be used in environmental monitoring and climate modeling to analyze changes in vegetation cover, water resources, and other environmental factors. In the field of wildfire recognition, detection, and segmentation, deep learning models have shown great potential\cite{ghali2023deep}. These models have been successfully applied to aerial and ground images to accurately classify wildfires, detect their presence, and segment the fire regions. Various deep learning architectures have been explored, including CNNs, one-stage detectors (such as YOLO), two-stage detectors (such as Faster R-CNN), and encoder-decoder models (such as U-Net and DeepLab). In the context of UAV images, a framework has been proposed for removing spatiotemporal objects from UAV images before generating the orthomosaic. The framework consists of two main processes: image segmentation and image inpainting. Image segmentation is performed using the Mask R-CNN algorithm, which detects and segments vehicles in the UAV images. The segmented areas are then masked to be removed. Image inpainting is carried out using the large mask inpainting (LaMa) method, a deep learning-based technique that reconstructs damaged or missing parts of an image \cite{park2022deep}. Additionally, a more extensive examination of aerial image-based FSIS can be found in the Table \ref{table:FSIS_aerial}. 

%%% Explaibility of FSIS

 Excitability in few-shot image segmentation, particularly in the context of remote sensing aerial images, have focused on the development of novel models and techniques that enhance the performance of segmentation tasks and provide insights into the decision-making process of the models. One such advancement is the Self-Enhanced Mixed Attention Network (SEMANet) proposed \cite{song2023self}. SEMANet utilizes three-modal (Visible-Depth-Thermal) images for few-shot semantic segmentation tasks. The model consists of a backbone network, a self-enhanced module (SE), and a mixed attention module (MA). The SE module enhances the features of each modality by amplifying the differences between foreground and background features and strengthening weak connections. The MA module fuses the three-modal features to obtain a better feature representation. Another advancement is the combination of a self-supervised background learner and contrastive representation learning to improve the performance of few-shot segmentation models \cite{cao2023few}. The self-supervised background learner learns latent background features by mining the features of non-target classes in the background. The contrastive representation learning component of the model aims to learn general features between categories by using contrastive learning. This approach has shown potential for enhancing the performance and generalization ability of few-shot segmentation models. Still, there are still some problems to solve in the field, such as how to deal with differences in performance caused by intra-class confusions and how to make models that are simple to understand and can be fairly accurate.  Future research should focus on the development of flexible few-shot object segmentation approaches that are capable of effectively handling lightweight models. These models should possess a higher level of interpretability for each of its components and demonstrate the ability to generalize across other domains.

\begin{table}
\caption{Overview of studies addressing challenges in few-shot image segmentation for remote sensing aerial images}
\scalebox{0.7}{
\begin{tabular}{|l|l|l|l|l|}
\hline
\textbf{References} & \textbf{Challenges} & \textbf{Dataset used} & \textbf{Year} \\
\hline

%%%%% few shot object detection 2021
SDM \cite{yao2021scale} & Broad direction for aligning query and support images  &  iSAID & 2021 \\
DMML-Net \cite{wang2021dmml} &  Metric learning to align query and support images   &  iSAID, DLRSD & 2021 \\

Chen et all. \cite{chen2022semi} &  Meta learning based alignment query and support images  &  Vaihingen and  Zurich Summer & 2022 \\ 

PQL \cite{wang2022queue} &  Prototype Queue learning based alignment query and support images  &  UDD and Vaihingen & 2022 \\ 

 Q2S \cite{puthumanaillam2023texture} &  Texture based alignment query and support images   & South Asia and Central Europe  & 2023 \\
FRINet \cite{cao2023few} &  Pair-to-pair matching network  &  iSAID & 2023 \\

%%%%%  2023

R2Net \cite{lang2023global} & global rectification and decoupled registration  &  iSAID & 2023 \\
PCNet \cite{lang2023progressive} & 'learning to learn' framework  &  iSAID & 2023 \\

%  MTGANs \cite{fan2023multitask} &  Robust of novel class  &  Sentinel-1, ERS-1/2, and GF-3 & 2023 \\ 
 
% Zhang et al. \cite{zhang2023text} & Multi-modal fusion for novel class   &  NWPU VHR and DIOR & 2023 \\
% TINet \cite{liu2023transformation} & Varying scale of novel class   &  NWPU VHR, DIOR, and HRRSD & 2023 \\
% BSFCDet \cite{liu2023transformation} & Generalization unseen class   &  SPPF-G, and SPP & 2023 \\

\hline
\end{tabular}}
\label{table:FSIS_aerial}
\end{table}

\subsubsection{Few-shot image segmentation benchmark datasets for remote sensing aerial images }

\begin{itemize}
    \item \textbf{iSAID} is a large-scale dataset for instance segmentation in aerial images. It contains 2,806 high-resolution images with annotations for 655,451 instances across 15 categories.

    \item \textbf{Vaihingen} consists of true orthophoto (TOP) images captured over the town of Vaihingen an der Enz, Germany. The images have a spatial resolution of 9 cm, which is quite high compared to many other aerial image datasets. The dataset also includes corresponding ground truth data, which provides pixel-wise annotations for six classes: impervious surfaces (such as roads and buildings), buildings, low vegetation (such as grass), trees, cars, and clutter/background.
    \item \textbf{DLRSD} contains images where the label data of each image is a segmentation image. This segmentation map is analyzed to extract the multi-label of the image. DLRSD has richer annotation information with 17 categories and corresponding label IDs.
\end{itemize}

\section{Discussions}
In this section, we aim to highlight interesting observations, common trends, and potential research gaps based on the in-depth analysis of the existing few-shot classification techniques across the three remote sensing data domains. The insights discussed in this section can serve as a guide for both current and future researchers in this field.

\begin{itemize}
\item Most of the methods described in the literature use different feature extraction models, with CNN-based models often serving as the backbone, as we've already talked about. Convolution-based few-shot learning models are still popular for classification tasks in all three domains. These models are capable of quickly adapting to new classes with few training examples, making them suitable for real-world applications. However, graph-based methods are becoming more popular for classifying SAR images, and they have only recently been used to classify VHR images. Graph-based methods are advantageous because they are able to capture the spatial relationships between objects, which is essential for classifying SAR and VHR images. Recently, vision transformer-based and incremental learning-based methods have emerged as alternatives for hyperspectral image classification. These methods have shown promise in achieving high accuracy with minimal training data, making them attractive for applications where labeled data is limited. 
\item The evaluation of the discussed works in hyperspectral image classification generally employs three commonly utilized metrics: overall accuracy (OA), average accuracy (AA), and kappa coefficient ($\kappa$). These metrics are frequently used to evaluate the classification performance of the proposed algorithms. In contrast, for VHR and SAR-based image classification, the classification accuracy (OA) is often utilized as the primary evaluation metric, although there are a few exceptions. Moreover, in most of the evaluation strategies adopted by the researchers, the proposed algorithms are run multiple times along with the state-of-the-art (SOTA) techniques, and the corresponding mean accuracy and its standard deviation are reported. This approach provides a more reliable and robust estimate of the classification performance, taking into account any potential variations in the results obtained across multiple runs.
\item In contrast to hyperspectral classification, it has been observed that there are currently few or no vision ViT-based few-shot classification methods proposed for SAR and VHR images. This could be attributed to the challenges associated with acquiring sufficient datasets for implementing effective and accurate ViT-based architectures for SAR images. Similarly, for VHR images, although there are existing models that use ViT-based classification, they are non-few-shot approaches such as the vanilla ViT-based model proposed by Zhang et al. \cite{zhang2021trs}. Consequently, there are considerable opportunities for researchers to explore the potential of few-shot ViT-based approaches for addressing the challenges associated with VHR remote sensing data classification.
\item The current state of research on few-shot classification approaches in the field of remote sensing does not seem to include much work on UAV or low-altitude aircraft-based images, as far as current knowledge suggests. This may be due to the unique nature of such images, which have been pointed out in previous studies such as \cite{gao2021fasts}. The differences in object sizes and perspectives, as well as the limited computational resources available for UAV-based operations, may be contributing factors to the scarcity of research in this area. In addition, the relatively smaller size of the UAV-based datasets may have posed challenges for few-shot learning methods, which often require a sufficiently large dataset to learn meaningful feature representations. However, with the increasing availability of UAV-based data, there may be opportunities for developing novel few-shot classification methods that can effectively leverage such data.

\item Furthermore, while few-shot learning has been studied extensively in the context of supervised classification, there is also potential for exploring its application in other remote sensing tasks such as unsupervised or semi-supervised learning, object detection, and semantic segmentation. Few-shot learning can provide an effective means of leveraging limited labeled data in these tasks, which can potentially lead to more accurate and efficient algorithms for remote sensing applications. Overall, while significant progress has been made in the application of few-shot learning to remote sensing data, there are still many research gaps and opportunities for further investigation. The exploration of new few-shot learning approaches, as well as the extension of existing methods to new applications and domains, can lead to more accurate and efficient algorithms for remote sensing tasks.

\item The utilization of XAI methodologies in conjunction with few-shot learning models for remote sensing applications can considerably enhance the interpretability of such models, thereby increasing their applicability in domains that are sensitive to potential risks. However, despite the significant promise held by XAI for few-shot learning in remote sensing, the current body of research in this field remains relatively nascent and further endeavors are necessary to fully realize its potential benefits.
\end{itemize}

\subsection{Computational considerations in few-shot learning}
Few-shot learning, as a niche within the broader domain of machine learning, warrants unique computational requirements. These requirements become particularly pertinent when the applications have real-time constraints. One of the most critical real-time applications lies in disaster monitoring using UAVs. The immediacy of feedback in such scenarios can drastically affect outcomes, emphasizing the significance of processing time.

Deep learning, which forms the foundation for many few-shot learning techniques, inherently demands high computational resources. Techniques such as CNNs are notorious for their computational intensity during both the training and inference phases. This computational cost can sometimes be a bottleneck, especially when rapid responses are essential. However, the evolving landscape of few-shot learning has seen the emergence of strategies aiming to mitigate these computational challenges:
\begin{itemize}
    \item Meta-learning, exemplified by approaches like MAML \cite{finn2017model}, offers an innovative solution. By optimizing model parameters to allow swift adaptation to novel tasks, these methods significantly reduce the computational overhead. This ensures that models can be fine-tuned efficiently, even when faced with new datasets.
    
    \item Wang et al.'s \cite{wang2022few} proposition of employing lightweight model architectures coupled with knowledge distillation techniques emerges as another viable strategy. By minimizing redundancies and unnecessary parameters, these models are streamlined to be more computationally efficient without compromising their predictive power.

    \item Graph-based methodologies, such as GraphSAGE \cite{yang2020hyperspectral}, and further extensions into GNN-based approaches \cite{yang2020learning}, provide alternatives to traditional CNNs. These methods, in certain dataset contexts, have demonstrated reduced computational complexity, making them attractive options.

\end{itemize}

Despite these advancements, it is noteworthy that a significant portion of few-shot learning methodologies has not been explicitly tailored for optimizing processing time. Recognizing this gap, future research could pivot towards crafting architectures specifically designed for real-time UAV applications. Several avenues could be pursued to enhance computational efficiency. These include embracing model compression techniques, such as pruning and quantization \cite{han2015deep}, leveraging efficient neural architecture search methods \cite{pham2018efficient}, and exploring hardware-software co-design strategies \cite{ham2021elsa} to fine-tune models for particular computational platforms. In all these endeavors, the overarching goal remains consistent: achieving rapid inference times without sacrificing model accuracy.

\section{Numerical experimentation of few-shot classification on UAV-based dataset}

\begin{table}
\caption{List of training, validation and test image sets for each class in our subset of the AIDER dataset. \label{tab6}}
\begin{tabular}{p{2cm}p{2cm}p{2cm}p{2cm}p{2cm}}
\hline
\textbf{Class} & \textbf{Train} & \textbf{Valid} & \textbf{Test} & \textbf{Total per Class}\\\hline

 Collapsed Building & 367 & 41 & 103 & 511\\
 Fire & 249 & 63 & 209 & 521\\
 Flood & 252 & 63 & 211 & 526\\
 Traffic & 232 & 59 & 194 & 485\\
Normal & 2107 & 527 & 1756 & 4390\\
\textbf{Total Per Set} & \textbf{3207} & \textbf{753} & \textbf{2473} & \textbf{6433}\\\hline

\end{tabular}
\end{table}

In order to address point 4 in the discussion section, a few-shot state-of-the-art (SOTA) method were employed to classify disaster scenes using the publicly available AIDER subset dataset. The evaluation involved the use of several few-shot methods such as the Siamese and Triplet Network, ProtoNet \cite{snell2017prototypical},  Relation Network \cite{sung2018learning}, Matching Network \cite{vinyals2016matching}, SimpleShot \cite{wang2019simpleshot}, TAsk-Dependent Adaptive Metric (TADAM) \cite{oreshkin2018tadam}, MAML \cite{finn2017model}, Meta-Transfer Learning (MTL) \cite{sun2019meta}, and Label Hallucination \cite{jian2022label}, which were originally proposed and evaluated in non-remote sensing datasets. The aim of the study was to evaluate the effectiveness of these methods in the remote sensing setting.  To compare the results obtained from such dataset against that of a satellite-based remote sensing image classification, we compared our findings with some of the methods utilized in the UC-Merced evaluation as performed by \cite{huang2021taes}; For the methods not listed there, we performed the simulation using the experimental condition as stipulated by \cite{huang2021taes}.

We conducted a 5-way 1-shot and 5-way 5-shot classification evaluation approach. The AIDER subset dataset consists of a total of 6433 images, classified into 5 categories, namely collapsed buildings, fires, floods, traffic accidents, and normal (non-disaster) classes, with 511, 521, 526, 485, and the rest of the images, respectively. The dataset subset is imbalanced, with more images in the normal class than in the other disaster classes, highlighting the potential benefits of few-shot learning approaches, as mentioned in previous sections. Table \ref{tab6} depicts the train-valid-test split ratio adapted for each class. All images are cropped to a pixel size of 224 $\times$ 224 and pre-processed by dividing each original pixel value by 255.  The learning rate for each algorithm is set as 0.001. ResNet12 is chosen as the feature extraction backbone for TADAM, ProtoNet, Matching Network, Relation Network, SimpleShot, MTL, and Label Hallucination.  For all of the methods, a common categorical cross-entropy loss is utilized, except for relation network, which utilized a mean-squared error loss. To tackle the problem of class imbalance, the training and validation samples were subjected to under-sampling by utilizing the RandomUnderSampler module, which was provided in the imblearn library package. All the simulations in this dataset were carried out for a total of 200 epochs using the Tensorflow Keras library in Python, and the Google Colab Pro+ platform with Tesla A100, V100, and T4 Graphical Processing Units (GPU) and Tensor Processing Units (TPU) were employed for computation. 

For the UC-Merced dataset, apart from the features as mentioned in section 4.2, 10 classes are utilized as the base training set, 5 classes are set aside as the validation set, and the remaining 6 classes are utilized as the novel test set. In line with \cite{huang2021taes}, the shapes of all images are cropped to 84 × 84 for feature extraction using their proposed feature encoder, with the momentum factor set to 0.1, and the learning rate set to 0.001. Due to all classes in UC-Merced having equal samples per class, methods to handle class imbalance are not needed. Once again, the common categorical cross-entropy is utilized as the loss function, except for relation network, which utilized a mean-squared error loss. 

Table \ref{tab7} presents the results of the simulations carried out on the AIDER subset using the few-shot evaluation approach mentioned earlier. Table \ref{tab8} the corresponding results on the UC-Merced dataset. The mean accuracy and standard deviation for 10 runs are reported for each method. For the Siamese and Triplet network, the results are only reported for the 5-way 1-shot evaluation, as only 1 pair of images is compared per episodic training (for the Triplet network, the anchor image is taken and compared with the positive vs the negative image each at a time, so 1 pair of images are still considered). It was observed that the mean accuracy for the 5-way 5-shot approach is generally higher than that of the 5-way 1-shot approach for all the methods utilized in the  two dataset, in agreement with the statement made earlier about the difficulty of few-shot learning with fewer shots. The Siamese network was found to outperform both Triplet and ProtoNet, demonstrating its effectiveness in feature extraction and embedding. Consistent with the trend observed in a previous study, MTL outperformed TADAM and ProtoNet in the AIDER  and the UC-Merced subset, while label hallucination yielded the highest performance with a metric value of over 81\% in the AIDER subset.

\begin{table}
\caption{The mean classification accuracy using the 5-way 1-shot and 5-way 5-shot learning evaluation on our AIDER subset simulation. Each method were ran for 10 times per setting. \label{tab7}}

\begin{tabular}{p{3cm}p{3cm}p{3cm}}
\hline
\textbf{Methods} & \textbf{5-way 1-shot} & \textbf{5-way 5-shot} \\
\hline
 Siamese Network & 75.0$\pm$0.97 & -\\
 Triplet Network & 65.3$\pm$0.77 & -\\   
 SimpleShot & 55.6$\pm$0.62 & 73.8$\pm$0.25\\
 TADAM & 69.0$\pm$0.33 & 76.4$\pm$0.55\\
 ProtoNet & 62.2$\pm$1.50 & 77.7$\pm$0.57\\
 Matching Network & 36.2$\pm$0.21 & 73.2$\pm$0.10\\
 Relation Network & 44.4$\pm$0.64 & 75.9$\pm$0.16\\
 MAML & 67.1$\pm$0.79 & 75.3$\pm$0.70\\
 MTL & 75.7$\pm$0.39 & 80.6$\pm$0.41\\
 Label Hallucination & 77.2$\pm$0.67 & 84.0$\pm$0.61\\\hline
\end{tabular}
\end{table}

\begin{table}
\caption{The mean classification accuracy using the 5-way 1-shot and 5-way 5-shot learning evaluation on UC-Merced. The Siamese, Triplet, SimpleShot, MTL, TADAM and Label Hallucination were ran for 10 times per setting, while for the remaining methods, the values were obtained from \cite{huang2021taes}. \label{tab8}}

\begin{tabular}{p{3cm}p{3cm}p{3cm}}
\hline
\textbf{Methods} & \textbf{5-way 1-shot} & \textbf{5-way 5-shot} \\
\hline
 Siamese Network & 63.8$\pm$0.94 & -\\
 Triplet Network & 61.2$\pm$0.92 & -\\
 SimpleShot & 57.8$\pm$0.77 & 66.1$\pm$0.34 \\
 TADAM & 63.4$\pm$0.67 & 73.0$\pm$0.72 \\
 ProtoNet & 52.6$\pm$0.70 & 65.9$\pm$0.57\\
 Matching Network & 46.2$\pm$0.71 & 66.7$\pm$0.56\\
 Relation Network & 48.9$\pm$0.73 & 64.1$\pm$0.54\\
 MAML & 43.7$\pm$0.68 & 58.4$\pm$0.64\\
 MTL & 65.8$\pm$0.36 & 73.4$\pm$0.74\\
 Label Hallucination & 67.3$\pm$0.82 & 75.3$\pm$0.73 \\\hline
\end{tabular}
\end{table}

\section{Explainable AI (XAI) in Remote Sensing}
XAI has become an increasingly crucial area of research and development in the field of remote sensing. As deep learning and other complex black-box models gain popularity for analysis of remote sensing data, there is a growing need to provide transparent and understandable explanations for how these models arrive at their predictions and decisions. Within remote sensing, explainability takes on heightened importance because model outputs often directly inform real-world actions with major consequences. For example, models identifying at-risk areas for natural disasters, pollution, or disease outbreaks can drive evacuations, remediation efforts, and public health interventions. If the reasoning behind these model outputs is unclear, stakeholders are less likely to trust and act upon the model's recommendations.

To address these concerns, XAI techniques in remote sensing aim to shed light inside the black box \cite{mohan2023quantitative}. Explanations can highlight which input features and patterns drive particular model outputs \cite{wang2019designing}. Visualizations can illustrate a model's step-by-step logic \cite{arrieta2020explainable}. Uncertainty estimates can convey when a model is likely to be incorrect or unreliable \cite{ling2015evaluation}. Prototypes and case studies have shown promise for increasing trust and adoptability of AI models for remote sensing applications ranging from climate monitoring to precision agriculture \cite{shaikh2022towards}. As remote sensors continue producing ever-larger and more complex datasets, the role of XAI will likely continue growing in importance. With thoughtful XAI implementations, developers can enable deep learning models to not only make accurate predictions from remote sensing data, but also provide the transparency and justifications required for stakeholders to confidently use these tools for critical real-world decision making. Recent approaches to XAI in the field of remote sensing are outlined below.

One notable development is the "What I Know" (WIK) method, which verifies the reliability of deep learning models by providing examples of similar instances from the training dataset to explain each new inference \cite{ishikawa2023example}. This technique demonstrates how the model arrived at its predictions.  

XAI has also been applied to track the spread of infectious diseases like COVID-19 using remote sensing data \cite{temenos2022novel}. By explaining disease prediction models, XAI enables greater trust and transparency. Additionally, XAI techniques have been used for climate adaptation monitoring in smart cities, where satellite imagery helps extract indicators of land use and environmental change \cite{sirmacek2022remote}. 

Several specific XAI methods show promise for remote sensing tasks, including Local Interpretable Model-Agnostic Explanations (LIME), SHapley Additive exPlanations (SHAP), and Gradient-weighted Class Activation Mapping (Grad-CAM)\cite{ishikawa2023example}. These methods highlight influential input features and image regions that led to a model's outputs. Grad-CAM produces visual heatmaps to indicate critical areas in an input image for each inference made by a convolutional neural network.

However, some challenges remain in fully integrating XAI into remote sensing frameworks. Practical difficulties exist in collecting labeled training data, extracting meaningful features, selecting appropriate models, ensuring generalization, and building reproducible and maintainable systems\cite{sirmacek2022remote}. There are also inherent uncertainties in modeling complex scientific processes like climate change that limit the interpretability of model predictions \cite{sirmacek2022remote}. Furthermore, the types of explanations provided by current XAI methods do not always match human modes of reasoning and explanation \cite{gevaert2022explainable}. Despite the challenges, XAI methods hold promise for enhancing few-shot learning approaches in remote sensing. Few-shot learning aims to learn new concepts from very few labeled examples, which is important in remote sensing where labeled data is scarce across the diversity of land cover types. However, the complexity of few-shot learning models makes their predictions difficult to interpret.

\subsection{XAI in Few-Shot Learning for Remote Sensing}
Most XAI methods for classification tasks are post-hoc, which cannot be incorporated into the model structure during training. Back-propagation \cite{chattopadhay2018grad,selvaraju2017grad,shrikumar2017learning,wang2020score} and perturbation-based methods \cite{schulz2020restricting} are commonly used in XAI for classification tasks. However, few works have been carried out on XAI for few-shot learning tasks. Initial work has explored techniques like attention maps and feature visualization to provide insights into few-shot model predictions for remote sensing tasks \cite{liu2022few}. Recently, a new type of XAI called SCOUTER \cite{li2021scouter} has been proposed, in which the self-attention mechanism are applied to the classifier. This method extracts discriminant attentions for each category in the training phase, allowing the classification results to be explainable. Such techniques can provide valuable insights into the decision-making process of few-shot classification models, increasing transparency and accountability, which is particularly important in remote sensing due to the high cost of acquiring and processing remote sensing data. In another recent work \cite{wang2022match}, a new approach to few-shot learning for image classification has been proposed that uses visual representations from a backbone model and weights generated by an explainable classifier. A minimum number of distinguishable features are incorporated into the weighted representations, and the visualized weights provide an informative hint for the few-shot learning process. Finally, a discriminator compares the representations of each pair of images in the support and query set, and pairs yielding the highest scores determined the classification results. This approach, when applied onto three mainstream datasets, achieved good accuracy and satisfactory explainability.

\subsection{Taxonomy of Explainable Few-Shot Learning Approaches}
Explainable few-shot learning techniques for remote sensing can be categorized along two main dimensions:
\subsubsection{Explainable Feature Extraction}
These methods aim to highlight influential features or inputs that drive the model's predictions.
\begin{itemize}
    \item Attention Mechanisms: Attention layers accentuate informative features and inputs by assigning context-specific relevance weights \cite{jetley2018learn}. They produce activation maps visualizing influential regions \cite{wang2022match,hong2021reinforced}. However, they don't explain overall reasoning process.
    \item Explainable Graph Neural Networks: Techniques like xGNNs \cite{yuan2020xgnn,moura2022explainable} can identify important nodes and relationships in graph-structured data. \cite{cheng2022attentive} puts forth attentive graph neural network modules that can provide visual and textual explanations illustrating which features are most crucial for few-shot learning. This provides feature-level transparency. But complete logic remains unclear.
    \item Concept Activation Visualization: Approaches like Grad-CAM produce saliency maps showing influential regions of input images \cite{selvaraju2016grad}. But local feature importance may not fully represent global decision process.
    \item Rotation-invariant Feature Extraction: The proposed rotation-invariant feature extraction framework in \cite{pintelas2023explainable} introduces an interpretable approach for extracting features invariant to rotations. This provides intrinsic visual properties rather than extraneous rotation variations.
\end{itemize}

\subsubsection{Explainable Decision Making}
These methods aim to directly elucidate the model's internal logic and reasoning.
\begin{itemize}
    \item Interpretable Models: Decision trees \cite{rudin2019stop} and rule lists \cite{letham2015interpretable} provide complete transparency into model logic in a simplified human-readable format. However, accuracy is often lower than complex models.
    \item Model-Agnostic Methods: Techniques like LIME \cite{ribeiro2016should} and SHAP approximate complex models locally using interpretable representations. But generating explanations can be slow at prediction time.
    \item Fairness Constraints: By imposing fairness constraints during training \cite{agarwal2018reductions} or transforming data into fair representations \cite{zemel2013learning}, biases can be mitigated. However, constraints may overly restrict useful patterns.
    \item Prototype Analysis: Analyzing prototypical examples from each class provides intuition into a model’s reasoning \cite{snell2017prototypical}. But limited to simpler instance-based models.
\end{itemize}

\begin{figure}
    \centering
    \includegraphics[scale=0.30]{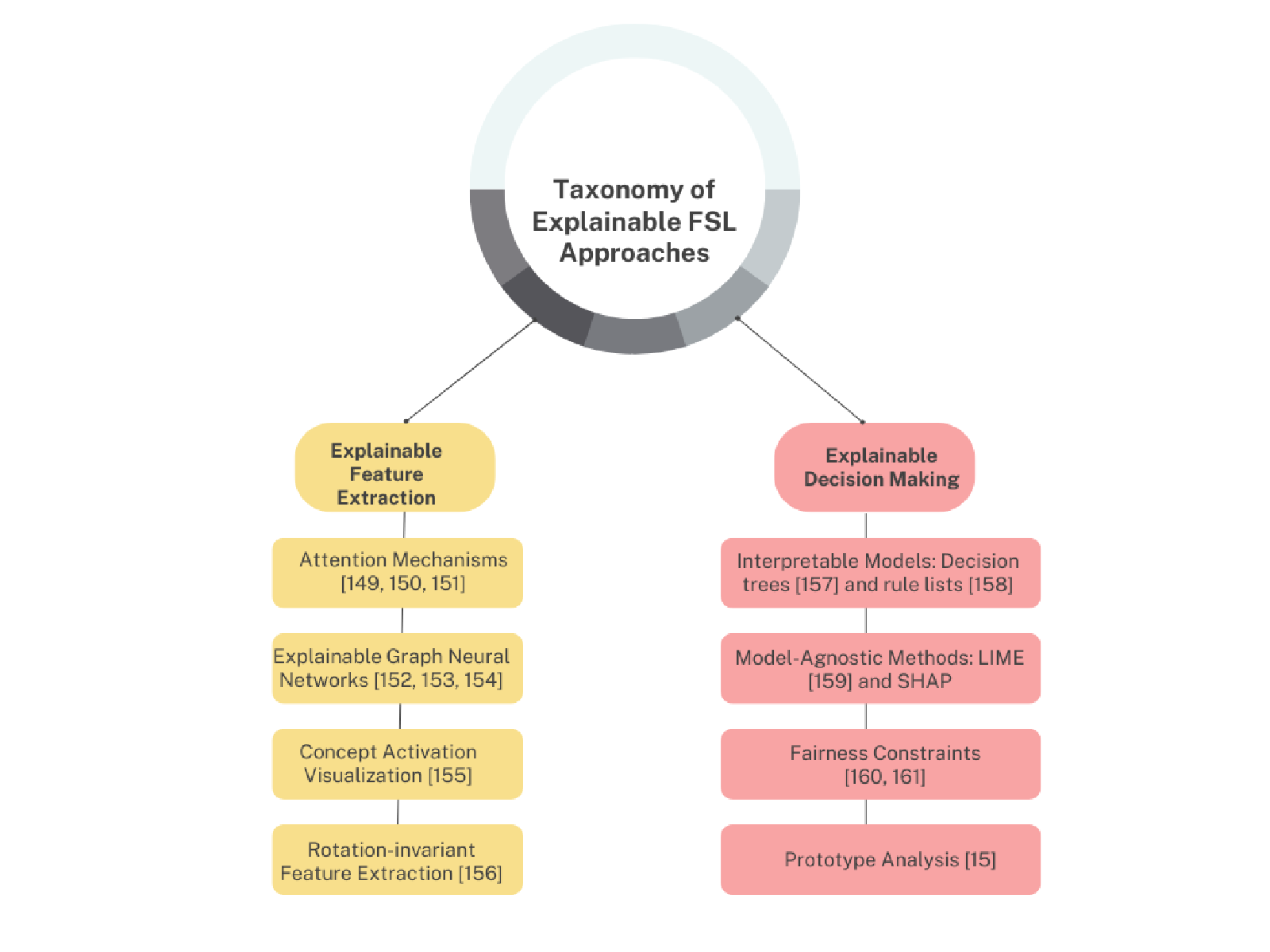}
    \caption{Taxonomy of explainable few-shot learning approaches}
    \label{fig:taxonomy_fsl}
\end{figure}

Overall, choosing suitable explainable few-shot learning techniques requires trading off accuracy, transparency, and efficiency based on the application requirements and constraints. A combination of feature and decision explanation methods is often necessary for complete interpretability. The taxonomy provides an initial guide to navigating this complex landscape of approaches in remote sensing contexts. For clarity, the taxonomy is also illustrated in Figure \ref{fig:taxonomy_fsl}.

\section{Conclusions and Future Directions}
In this comprehensive review, we provided a comprehensive analysis of recent few-shot learning techniques for remote sensing across various data types and platforms. Compared to previous reviews \cite{sun2021research}, we expanded the scope to include UAV-based datasets. Our quantitative experiments demonstrated the potential of few-shot methods on various remote sensing datasets. We also emphasized the growing importance of XAI to increase model transparency and trustworthiness.

While progress has been made, ample opportunities remain to advance few-shot learning for remote sensing. Future research could explore tailored few-shot approaches for UAV data that account for unique image characteristics and onboard computational constraints. Vision transformer architectures could also be investigated for few-shot classification of very high-resolution remote sensing data. A key challenge is reducing the performance discrepancy between aerial and satellite platforms. Developing flexible techniques that handle diverse data effectively is an open problem that warrants further investigation.

On the XAI front, further work is needed to address issues unique to remote sensing like scarce labeled data, complex earth systems, and integrating domain knowledge into models. Techniques tailored for few-shot learning specifically could benefit from more research into explainable feature extraction and decision making. Explainability methods that provide feature-level and decision-level transparency without sacrificing too much accuracy or efficiency are needed. There is also potential to apply few-shot learning and XAI to new remote sensing problems like object detection, semantic segmentation, and anomaly monitoring.

To end, few-shot learning shows increasing promise for efficient and accurate analysis of remote sensing data at scale. Integrating XAI can further improve model transparency, trust, and adoption by providing human-understandable explanations. While progress has been made, ample challenges and opportunities remain to realize the full potential of few-shot learning and XAI across the diverse and rapidly evolving remote sensing application landscape. Advances in these interconnected fields can pave the way for remote sensing systems that learn quickly from limited data while remaining transparent, accountable, and fair.

\section*{Acknowledgements}
This research/project is supported by the Civil Aviation Authority of Singapore and Nanyang Technological University, Singapore under their collaboration in the Air Traffic Management Research Institute. Any opinions, findings and conclusions or recommendations expressed in this material are those of the author(s) and do not reflect the views of the Civil Aviation Authority of Singapore.

\section*{Declarations}

The authors have no conflicts of interest to declare.

\bibliography{ref_xai_few_shot}% common bib file
%% if required, the content of .bbl file can be included here once bbl is generated
%%\input sn-article.bbl

%% Default %%
%%\input sn-sample-bib.tex%

\end{document}